%% file: draft.tex
\documentclass[twocolumn,showpacs,preprintnumbers,amsmath,amssymb,amsfonts,final,a4paper,aps,citeautoscript,footinbib,prb,superscriptaddress]{revtex4-1}
\usepackage{times}
\usepackage[pdftex]{graphicx,color}    
\usepackage{epstopdf}
\usepackage[colorlinks,bookmarks=true,citecolor=blue,linkcolor=blue,urlcolor=blue, breaklinks=true]{hyperref}
\usepackage{simplewick}
\usepackage[latin1]{inputenc}
\usepackage{mathrsfs}
\usepackage[vcentermath]{youngtab}
\newcommand{\bolN}{\boldsymbol{N}}
\newcommand{\be}{\text{e}}
\usepackage{epstopdf}
\begin{document}
\title{Symmetry-protected topological phases in two-leg SU($N$) spin ladder with unequal spins}
\author{S. Capponi} 
\affiliation{Laboratoire de Physique Th\'eorique, IRSAMC, 
  Universit\'e de Toulouse, CNRS, UPS, France.}
\author{P. Fromholz} 
\affiliation{The Abdus Salam International Centre for Theoretical Physics, Strada Costiera 11, 34151 Trieste, Italy}
\affiliation{SISSA, via Bonomea 265, 34136 Trieste, Italy}
\author{P. Lecheminant} 
\affiliation{Laboratoire de Physique Th\'eorique et
 Mod\'elisation, CNRS, Universit\'e de Cergy-Pontoise,
F-95300 Cergy-Pontoise Cedex, France.}
\author{K. Totsuka} 
\affiliation{Yukawa Institute for Theoretical Physics, 
Kyoto University, Kitashirakawa Oiwake-Cho, Kyoto 606-8502, Japan.}
\date{\today}
\pacs{{75.10.Pq} 
}
\begin{abstract}
Chiral Haldane phases are examples of one-dimensional topological states of matter which 
are protected by projective SU($N$) group (or its subgroup $\mathbb{Z}_N \times \mathbb{Z}_N$) with $N>2$.  
The unique feature of these symmetry protected topological (SPT) phases is that they are accompanied by 
inversion-symmetry breaking and the emergence of different left and right edge states which transform, for instance, 
respectively in the fundamental ($\bolN$) and anti-fundamental ($\overline{\bolN}$) representations of SU($N$). 
We show, by means of complementary analytical
and numerical approaches, that these chiral SPT phases as well as the non-chiral ones are realized as the ground states 
of a generalized two-leg SU($N$) spin ladder in which the spins in the first chain transform in $\bolN$ 
and the second in $\overline{\bolN}$.  In particular, we map out the phase diagram for $N=3$ and $4$ to show 
that {\em all} the possible symmetry-protected topological phases with projective SU($N$)-symmetry appear 
in this simple ladder model.  
\end{abstract}
\maketitle
\section{Introduction}
\label{sec:intro}
Spin ladders have been a constant source of interest since the original motivation that 
these systems are intermediates between one and two dimensions 
which could shed light on spin liquid physics in higher dimensions. \cite{Dagotto-R-96}
The simplest of them is two spin-1/2 antiferromagnetic Heisenberg chains (with spin-exchange $J_{\parallel}>0$)  
which are coupled to each other by an interchain coupling $J_{\perp}$.  
In the absence of this interchain coupling, the Heisenberg spin chains have a gapless energy spectrum 
with spin-spin correlations functions displaying universal power-law behavior 
(up to logarithmic corrections).\cite{Gogolin-N-T-book,Giamarchi-book-04}
The massless elementary excitations are the spinons with fractional spin-1/2 quantum numbers and 
appear only in pairs in any physical states with integer total spin. \cite{Faddeev-T-81}
In this respect, the standard magnon-excitations carrying spin-1 quantum number are deconfined into two spinons 
and this leads to an incoherent background in the dynamical spin susceptibility measured 
in neutron-scattering experiments. \cite{Mourigal-E-K-C-S-R-13}

However, these massless spinon excitations become confined into massive spin-1 triplons as soon as an infinitesimal coupling $J_{\perp}$ 
is introduced between the chains and a two-leg spin ladder provides a fundamental system 
to investigate the confinement of fractional quantum number particles. \cite{Shelton-N-T-96,Lake-etal-10}
Various gapful quantum phases have been predicted both analytically and numerically for two-leg spin ladders with 
a spin-exchange $J_{\times}$  
between the diagonals of the ladder \cite{Solyom-T-86,Schulz-86,Xian-95,White-96,Kim-F-S-S-00,Allen-E-N-00,Starykh-B-04,Kim-L-S-08,Hikihara-S-10}, 
a four-spin exchange interaction  
\cite{Nersesyan-T-97,Lauchli-S-T-03,Momoi-H-N-H-03,Gritsev-N-B-04,Lecheminant-T-05,Lecheminant-T-06-SU4},  
and a zigzag coupling \cite{White-A-96,Allen-S-97,Nersesyan-G-E-98,Lavarelo-R-14}.  
One interesting phase, found in these spin ladders,  is a non-degenerate fully gapped phase adiabatically 
connected to the Haldane phase of the spin-1 antiferromagnetic Heisenberg chain. \cite{Haldane-PLA-83,Haldane-PRL-83}
The latter is the paradigmatic example of one-dimensional (1D) symmetry-protected topological (SPT) phases 
with spin-1/2 edge states protected by, e.g., (on-site) $\mathbb{Z}_2 \times \mathbb{Z}_2$ symmetry. 
\cite{Gu-W-09,Pollmann-T-B-O-10,Pollmann-B-T-O-12}
The emergence of this SPT in two-leg spin ladder systems can be simply understood by considering 
the limit of large ferromagnetic interchain coupling $J_{\perp} \rightarrow - \infty$.   
In this limit, the strong $J_{\perp}$ favors the triplet states over the singlet one on each rung 
so that the low-energy effective Hamiltonian is given by the spin-1 Heisenberg chain.  
The resulting Haldane physics is adiabatically connected to the weak-interchain limit where the
low-energy physics is described in terms of four massive Majorana fermions 
with spin-1/2 edge states.\cite{Totsuka-S-ladder-95,Shelton-N-T-96,Lecheminant-O-02}

In this paper, we generalize this idea of stacking Heisenberg spin chains 
to realize various 1D gapped (SPT, in particular) phases to higher continuous symmetry group SU($N$) ($N \geq 3$).  
The 1D bosonic SPT phases with on-site (protecting) symmetry group $G$ 
are known to be classified by the second cohomology group ${\cal H}^{2}(G, \text{U(1)})$.\cite{Chen-G-W-11,Schuch-G-C-11,Fidkowski-K-11,Chen-G-L-W-13} 
When $G$ is the projective unitary group $\text{PSU($N$)} \simeq \text{SU($N$)}/\mathbb{Z}_N $, 
${\cal H}^{2}(\text{PSU($N$)}, \text{U(1)}) = \mathbb{Z}_N$ and $N-1$ non-trivial SPT phases 
are expected,\cite{Duivenvoorden-Q-ZnxZn-13,Else-B-D-13}   
while in the case of spin-rotational invariance, i.e., $G=$ SO(3) $\simeq \text{SU(2)}/\mathbb{Z}_2 $, there is a single non-trivial SPT phase. 
For practical purposes, it is convenient to label these phases by a pair of projective representations 
$(\mathcal{R}_{\text{e}},\overline{\mathcal{R}}_{\text{e}})$ under which the left and right edge states transform.  
One interesting class of these SPT phases is the so-called {\em chiral} SPT phase 
$(\mathcal{R}_{\text{e}},\overline{\mathcal{R}}_{\text{e}})=(\bolN, \overline{\bolN})$, 
which is a non-degenerate fully gapped phase  where the left (right) edge state belongs to the fundamental $\bolN$ 
(its conjugate $\overline{\bolN}$) 
representation of the SU($N$) group.\cite{Affleck-K-L-T-88,Rachel-S-S-T-G-10,Katsura-H-K-08,Morimoto-U-M-F-14,Roy-Q-18}
The phase, which is protected by on-site $\text{PSU($N$)}$ symmetry or its subgroup $\mathbb{Z}_N \times \mathbb{Z}_N$,  
is featureless but breaks the inversion or site-parity symmetry since the left and right boundary spins are different 
(i.e., $\mathcal{R}_{\text{e}} \neq \overline{\mathcal{R}}_{\text{e}}$) and related by conjugation. \cite{Morimoto-U-M-F-14}  
These states are also interesting in the light of 
a recent proposal\cite{Wang-A-R-18} to use the two degenerate chiral SPT states with broken parity 
to realize a stable topological qubit ({\em valence-bond-solid qubit}).  

For $\text{SU($N \geq 3$)}$, the two fundamental representations $\boldsymbol{N}$ and $\overline{\boldsymbol{N}}$, 
which are SU($N$)-analogues of spin-1/2, are not identical.  
Therefore, we can think of two different generalizations of two-leg spin-1/2 ladders to SU($N$).  
Clearly, an obvious generalization is to just replace the two spin-1/2 chains with two {\em identical} 
SU($N$) spin chains in the defining representation $\boldsymbol{N}$.  
However, the Lieb-Schultz-Mattis-type argument\cite{Affleck-L-86}
\footnote{It is straightforward to adapt the argument in Ref.~\onlinecite{Affleck-L-86} for the $n_{\text{leg}}$-leg ladder to show that SPT phases are excluded unless $\sum_{j=1}^{n_{\text{leg}}}n_{\text{Y}}(j)$ [$n_{\text{Y}}(j)$ being the number of boxes in the Young diagram characterizing the SU($N$) spins of the $j$-th leg] is divisible by $N$. For the two-leg ladder made of identical chains (in $\boldsymbol{N}$),  $\sum_{j=1}^{n_{\text{leg}}}n_{\text{Y}}(j)=2$ while $\sum_{j=1}^{n_{\text{leg}}}n_{\text{Y}}(j)=N$ for the two-leg ladder \eqref{eqn:def-2leg-ladder}.}
and extensive analytical/numerical studies\cite{vdBossche-A-L-M-01,Lecheminant-T-15,Weichselbaum-C-L-T-L-18} 
convincingly rule out SPT phases in such ladders with $N\geq 3$.  

\begin{figure}[ht]
\begin{center}
\includegraphics[scale=0.8]{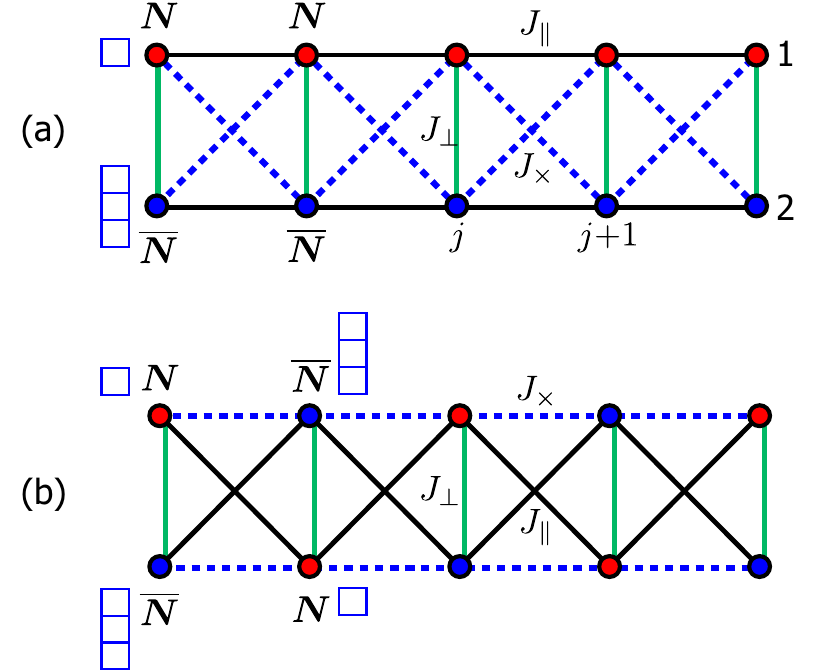}
\caption{
(a) $\boldsymbol{N}$-$\overline{\boldsymbol{N}}$ two-leg ladder given by Eq.~\eqref{eqn:def-2leg-ladder} 
and (b) its equivalent representation.}
\label{fig:two-ladders}
\end{center}
\end{figure}

Alternatively, we could put spins in the conjugate representation $\overline{\boldsymbol{N}}$ on the second chain 
and couple two {\em different} chains [see Fig.~\ref{fig:two-ladders}(a)].  
Specifically, we investigate the zero-temperature phase diagram of the following two-leg SU($N$) spin ladder 
with standard interchain coupling ($J_{\perp}$) as well as an interaction along the diagonals ($J_{\times}$):
\begin{equation}
\begin{split}
{\cal H}_{\bolN \text{-}\overline{\bolN}} =&  J_{\parallel} \sum_i \left( S^{A}_{1,i}  S^{A}_{1,i+1} 
+ \overline{S}^{A}_{2,i}  \overline{S}^{A}_{2,i+1} \right)
+   J_{\perp} \sum_i  S^{A}_{1,i}  \overline{S}^{A}_{2,i}    \\
&+ J_{\times} \sum_i \left( S^{A}_{1,i}  \overline{S}^{A}_{2,i+1}  + \overline{S}^{A}_{2,i}S^{A}_{1,i+1} \right) ,
\end{split}
\label{eqn:def-2leg-ladder}
\end{equation}
where $S^{A}_{1,i}$ and $\overline{S}^{A}_{2,i}$ ($A=1, \ldots, N^2 -1$) denote the SU($N$) spin operators 
on the $i$ th site on the chains 1 and 2 
which transform in the fundamental representation $\bolN$ of SU($N$) and its conjugate $\overline{\bolN}$, respectively.  
These spin operators are normalized as $\text{Tr}(S^{A} S^{B})=\text{Tr}(\overline{S}^{A} \overline{S}^{B})=\delta^{AB}/2$.   
By combined use of analytical approaches (weak-coupling low-energy field theories, large-$N$ expansion, etc.) and numerical 
density-matrix renormalization group (DMRG) calculations\cite{White-92,White-H-93} 
(see, e.g., Ref.~\onlinecite{Schollwoeck-11} for a review),  
we shall show below that the two-leg spin ladder \eqref{eqn:def-2leg-ladder} with {\em unequal} SU($N$) representations
exhibits a very rich phase structure. 
In particular, we argue that the simple enough ladder model \eqref{eqn:def-2leg-ladder} realizes {\em all} possible SPT phases 
protected by on-site PSU($N$) symmetry, including the ones recently found in the theoretical studies 
of SU($N$) ultracold fermions in a 1D optical lattice.\cite{Nonne-M-C-L-T-13,Bois-C-L-M-T-15,Fromholz-C-L-P-T-19} 
Recently, the Hamiltonian \eqref{eqn:def-2leg-ladder} has been investigated 
in the limit $J_{\perp} \to - \infty$ in Ref.~\onlinecite{Ueda-M-M-18} leading to some interesting observations.  

Another physical motivation to study this model is to understand the mechanism of the chiral SPT phases recently found numerically 
in the study of SU($N$) (with $N$-odd) fermions trapped in a 1D array of optical double wells.\cite{Fromholz-C-L-P-T-19}
In fact, the model \eqref{eqn:def-2leg-ladder} 
appears, in the limit of strong on-site interactions, as a skeletonized effective Hamiltonian of the original fermion model 
(see Sec.~II B in Supplementary Materials of Ref.~\onlinecite{Fromholz-C-L-P-T-19} for more details). 
Therefore, the study of the model \eqref{eqn:def-2leg-ladder} would help us to understand how chiral SPT phases with spontaneously 
broken parity symmetry arise in the Mott-insulating phase of the fermion model. 

Some remarks are in order about useful limits of the model \eqref{eqn:def-2leg-ladder}.  
In the absence of the interchain couplings ($J_{\perp}, J_{\times} = 0$), the model \eqref{eqn:def-2leg-ladder}  
 becomes two decoupled SU($N$) Sutherland models\cite{Sutherland-75} 
 in which one of the two chains is in the fundamental representation ($\boldsymbol{N}$) 
 and the other in its conjugate ($\overline{\boldsymbol{N}}$).   
 Both models are exactly solvable by means of the Bethe ansatz\cite{Sutherland-75} 
 and display a quantum critical behavior in the SU($N$)$_1$ universality class 
 with central charge $c=N-1$. \cite{Sutherland-75,Affleck-NP86,Affleck-88,Fuhringer-R-T-G-S-08,James-K-L-R-T-18,Nataf-M-18}  
 As will be discussed in Sec.~\ref{sec:strongcoupling}, we expect several different phases in the strong-coupling limit $|J_{\perp}|  \gg |J_{\parallel}|,|J_{\times}|$; when $J_{\perp}  >0$, the system is 
in a non-degenerate gapped singlet phase which is the generalization of the rung-dimer (RD) phase 
of the two-leg SU(2) spin ladder\cite{Dagotto-R-96,Gogolin-N-T-book,Giamarchi-book-04} to general $N$. 
When $J_{\perp} < 0$,  the model \eqref{eqn:def-2leg-ladder} reduces to a non-trivial SU($N$) spin chain in the adjoint representation 
which is expected to host several distinct phases.  Preliminary investigation suggested that there are various interesting
competing phases (including topological ones) in the positive $J_\times$ region. Therefore, we will
consider mostly the case with $J_\times>0$ in what follows and leave the other cases for future work.

The line $J_{\times} =J_{\parallel}$ is special in that the spin ladder \eqref{eqn:def-2leg-ladder} 
becomes the composite-spin representation of a single Heisenberg {\em chain} with the local SU($N$) spin operators defined by: 
$G^{A}_i = S^{A}_{1,i}  + \overline{S}^{A}_{2,i} $. Since the total SU($N$) ``spin'' of each rung [i.e., the SU($N$) representation living on the rung which is either in singlet or adjoint; see Eq.~\eqref{eqn:CG-N-Nbar}] is conserved when $J_{\times} =J_{\parallel}$, 
the entire Hilbert space splits into disjoint sectors corresponding to sets of SU($N$) representations on the individual rungs.   
When all the rungs are in the adjoint representation (this is the case e.g., when $J_{\perp}$ is negatively large), 
then the SU($N$) Heisenberg spin chain in the adjoint representation 
is obtained.  The ground state is known\cite{Rachel-S-S-T-G-10,Morimoto-U-M-F-14} 
to be in a fully gapped translational invariant phase with spontaneously broken inversion symmetry 
dubbed the chiral Haldane phase with the emergent edge states $(\bolN,\overline{\bolN})$ or  $(\overline{\bolN},\bolN)$  
related by the inversion symmetry.  

Last, in the limit of large $N$, a substantial simplification occurs and the ground state is understood by 
the energetics of valence-bond patterns (see Sec.~\ref{sec:large-N}).  We shall use this property to investigate 
the phase structure of the model \eqref{eqn:def-2leg-ladder} in the large-$N$ limit. 

The paper is organized as follows. In Sec.~\ref{sec:strongcoupling}, we discuss the strong-coupling regions 
$|J_{\parallel}/J_{\perp}| , |J_{\times}/J_{\perp}| \to 0$ of the two-leg spin ladder \eqref{eqn:def-2leg-ladder}.  
In particular, for ferromagnetically large $J_{\perp}$, we derive an effective extended Heisenberg model 
in the adjoint representation that contains a new interaction present only for $N \geq 3$.  The competition between 
the ordinary Heisenberg interaction and the additional one determines the phase structure in the strong-coupling region. 
The large-$N$ limit, in which the determination of the ground state reduces to simple energetics of valence-bond states, 
is considered in Sec.~\ref{sec:large-N}.  There we find three dominant phases one of which is thought of as 
a large-$N$ analogue of the chiral SPT phases.  
The low-energy approach appropriate in the weak-coupling ($J_{\parallel}>0$, $|J_{\perp}|,|J_{\times}|/ J_{\parallel} \ll 1$) 
regions  is presented in Sec.~\ref{sec:weakcoupling}
and we argue that the low-energy physics is described by an SU($N$)$_2$ and a ${\mathbb Z}_N$ parafermionic conformal field theories (CFT).  \cite{DiFrancesco-M-S-book}
Applying the semi-classical approximation, we find a translationally-invariant phase with spontaneously broken inversion symmetry 
which we identify as the chiral SPT phase.  
Unfortunately, the edge states, which give the fingerprint of the SPT phases, cannot be explored within our approach 
unlike in the $N=2$ case\cite{Lecheminant-O-02} due to the lack of free-field description of the underlying CFTs. 
The results of extensive DMRG calculations in the $N=3$ and $N=4$ cases are described in Sec.~\ref{sec:dmrg} 
where we map out the phase diagram of the generalized SU($N$) two-leg spin ladder \eqref{eqn:def-2leg-ladder} 
with the help of various local and non-local order parameters as well as the entanglement spectrum.   
The phase diagrams presented in Figs.~\ref{fig:dmrg_su3} and \ref{fig:dmrg_su4} are the central results of this paper.  
By explicitly calculating local spin densities, 
we also correctly identify the edge states expected for the two chiral SPT phases.  
Finally, a summary of the main results is given in Sec.~\ref{sec:conclusion} and some technical details are provided in the Appendices.

\section{Strong-coupling limit}
\label{sec:strongcoupling}
The strong-coupling expansion starts from the case of isolated rungs, i.e., $J_{\parallel}=J_{\times}=0$.  
Then, the Clebsch-Gordan decomposition 
\begin{equation}
\bolN \times \overline{\bolN} \simeq \mathbf{1} \oplus \mathbf{(\bolN^2-1)} 
\label{eqn:CG-N-Nbar}
\end{equation} 
tells that the $N^2$ states on each rung are decomposed into the singlet 
and the $(N^2-1)$-dimensional adjoint representations.    
Accordingly, the energy of each rung is given by
\begin{equation}
J_{\perp} S^{A}_{1,i} \overline{S}^{A}_{2,i}  = 
\begin{cases}
-J_{\perp} (N^2-1)/(2N)  & \text{singlet} \\
J_{\perp} /(2N) & \text{adjoint} \; .
\end{cases}
\label{eqn:SdotS-N-Nbar}
\end{equation}
Therefore, the strong-coupling region with $J_{\perp}>0$ is in the non-degenerate RD phase analogous to 
that found in the $N=2$ case;\cite{Dagotto-R-96} the ground state is a product of local SU($N$)-singlets 
on the individual rungs and 
the action of the SU($N$) generators create $(N^2-1)$-plet of gapful excitations [an SU($N$) generalization of the triplon] 
over this ``non-magnetic'' ground state.  

In the case of large negative $J_{\perp}$, where $\bolN$ and $\overline{\bolN}$ form the $(N^2-1)$-dimensional adjoint 
representation,  the strong-coupling limit is even less trivial.    One may naively expect that the SU($N$) Heisenberg chain 
for the preformed SU($N$) spins on the individual rungs is obtained as in the usual $S=1/2$ two-leg ladder.  Below, we will show that the resulting effective Hamiltonian contains an extra interaction that does not exist in the SU(2) case 
(on the other hand, the above expectation is correct 
in the case of the $\bolN$-$\bolN$ ladder which is symmetric in the upper and lower chains; 
see Appendix \ref{sec:N-N-ladder-strong-coupling}).   

We start from the $(N^{2}-1)$-fold degenerate ground states on each rung: 
\begin{equation}
|A\rangle = \sqrt{2} \sum_{\alpha,\beta=1}^{N} [S^A]_{\alpha\beta} |\alpha\rangle\otimes |\bar{\beta}\rangle \quad 
(A=1,\ldots, N^{2}-1) \; ,   
\label{eqn:adj-states-by-3-3bar}
\end{equation}
where $\{S^{A}\}$ are the hermitian SU($N$) generators in the $N$-dimensional defining representation ($\bolN$) 
satisfying $\text{Tr}(S^{A} S^{B})=\delta^{AB}/2$ (a possible choice of such generators is given in 
Appendix \ref{eqn:ZnxZn-gen-by-spins}), and $|\alpha\rangle$ ($|\bar{\alpha}\rangle$) 
are the states in $\bolN$ ($\overline{\bolN}$).  
It is easy to check that these are orthonormal:
\begin{equation}
\begin{split}
\langle A | B \rangle 
& = 2 \sum_{\alpha,\beta=1}^{N} \sum_{\mu,\nu=1}^{N} [S^A]_{\mu\nu}^{\ast} [S^B]_{\alpha\beta} 
\langle \mu | \otimes \langle \bar{\nu}| \alpha\rangle\otimes | \bar{\beta}\rangle   \\
&= 2\sum_{\alpha,\beta=1}^{N} [S^A]^{\ast}_{\alpha\beta} [S^B]_{\alpha\beta} 
= 2\, \text{Tr}(S^{A}S^{B}) 
=\delta^{AB} \; .
\end{split}
\end{equation}
The products of $\otimes_{j} |A_{j}\rangle_{j}$ on the individual rungs form a degenerate manifold of 
the strong-coupling ground states on which the Hamiltonian \eqref{eqn:def-2leg-ladder} acts non-trivially.   

To carry out degenerate perturbation theory, 
let us calculate the matrix elements of the SU($N$) generators with respect to $\{ |A\rangle \}$.  
We begin by the generators in $\bolN$ which act on the states of the first chain 
[i.e., on the $\bolN$ part of \eqref{eqn:adj-states-by-3-3bar}]: 
\begin{equation}
\hat{S}^{A} = \sum_{\alpha,\beta=1}^{N} [S^A]_{\alpha\beta} |\alpha\rangle\langle \beta | \; .
\end{equation}
Using Eq.~\eqref{eqn:adj-states-by-3-3bar}, we can readily evaluate
\begin{equation}
\begin{split}
\langle C | \hat{S}^{A} | B \rangle 
&= 2 \, \text{Tr} (S^B S^C S^A) 
= 2 \, \text{Tr}(S^C S^A S^B)    
\; ,
\end{split}
\label{eqn:G_A-in-3}
\end{equation}
where we have used the hermiticity of $\{ S^{A} \}$.  
Similarly, for the generators in the conjugate representation $\overline{\mathbf{N}}$
\begin{equation}
\hat{\overline{S}}{}^{A} = \sum_{\alpha,\beta=1}^{N} [-S^{A}]^{\text{T}}_{\alpha\beta} 
|\bar{\alpha}\rangle\langle \bar{\beta} | \; ,
\end{equation}
we obtain the following:
\begin{equation}
\langle C | \hat{\overline{S}}{}^{A} | B \rangle 
= - 2 \, \text{Tr} (S^C S^B S^A) 
= - 2 \, \text{Tr}(S^B S^A S^C)  \; .
\label{eqn:G_A-in-3bar}
\end{equation}
Now we want simplify the expressions \eqref{eqn:G_A-in-3} and \eqref{eqn:G_A-in-3bar}.  
First, we note that, for $\boldsymbol{N}$, we have the following commutation and anti-commutation relations:
\begin{equation}
\begin{split}
&  [ S^A , S^B] = i f_{ABC} S^{C}  \; ,  \\
&  \{ S^A , S^B \} =  d_{ABC} S^C + \frac{1}{N} \delta_{AB} \mathbf{1}  \; ,
\end{split}
\label{eqn:comm-rel-defining-rep}
\end{equation}
where $\{ f_{ABC} \}$ are the usual completely antisymmetric structure constants, while  
$\{ d_{ABC} \}$ are the {\em symmetric} structure constants which do not exist in SU(2).  
Combining them, we see that the product $S^A S^B$ can be written as:
\begin{equation}
S^A S^B = \frac{1}{2}  i f_{ABC} S^{C} + \frac{1}{2} d_{ABC} S^C  
+ \frac{1}{2N} \delta_{AB} \mathbf{1} \; ,
\end{equation}
which enables us to rewrite \eqref{eqn:G_A-in-3} and \eqref{eqn:G_A-in-3bar} as:
\begin{equation}
\begin{split}
& \langle C | \hat{S}^{A} | B \rangle =  2 \, \text{Tr}(S^C S^A S^B)  
= \frac{1}{2}  i f_{ABC} + \frac{1}{2} d_{ABC}  
 \\
& \langle C | \hat{\overline{S}}{}^{A} | B \rangle =  - 2 \, \text{Tr}(S^B S^A S^C)   
= \frac{1}{2}  i f_{ABC} - \frac{1}{2} d_{ABC} 
\; .
\end{split}
\end{equation}
Now it is convenient to introduce another set of $(N^2-1)\times(N^2-1)$-dimensional matrices 
$\{\mathcal{D}^A \}$ acting on the adjoint representation:
\begin{equation}
[\mathcal{D}^A]_{CB} \equiv  d_{ABC} 
\quad (A=1,\cdots, N^2 -1) \;  .
\end{equation} 
Then, we have:
\begin{equation}
\begin{split}
& \langle C | ( \hat{S}^{A} + \hat{\overline{S}}{}^{A}) | B \rangle = 
 i f_{ABC} \equiv  [\mathcal{S}^A]_{CB}   \\
& \langle C | ( \hat{S}^{A} - \hat{\overline{S}}{}^{A}) | B \rangle = 
 d_{ABC} = [\mathcal{D}^A]_{CB}   \; ,
\end{split}
\label{eqn:matrix-elem-rung}
\end{equation} 
where $\{\mathcal{S}^A\}$ are the SU($N$) generators in the adjoint representation.  
The point is that when a pair of spins are made up of different representations $\bolN$ and $\overline{\bolN}$, 
the action of the SU($N$) generators $\hat{S}{}^{A}$ and $\hat{\overline{S}}{}^{A}$ 
on the adjoint module contains not only $\mathcal{S}^A$ but also 
the new matrices $\mathcal{D}^A$ [note that, in the case of SU(2), 
$\langle C | ( \hat{S}_{A} - \hat{\overline{S}}_{A}) | B \rangle =0$ 
identically and operators like $\mathcal{D}^{A}$ never appear.].   

Now we are ready to calculate the first-order perturbation.  
First, we note:
\begin{equation}
\begin{split}
& J_{\parallel} \left\{ \hat{S}{}_{1,i}^{A}  \hat{S}{}_{1,i+1}^{A} + 
\hat{\overline{S}}{}_{2,i}^{A}  \hat{\overline{S}}{}_{2,i+1}^{A}  \right\}    
+ J_{\times}  \left\{ \hat{S}{}_{1,i}^{A} \hat{\overline{S}}{}_{2,i+1}^{A} 
+  \hat{\overline{S}}{}_{2,i}^{A} \hat{S}{}_{1,i+1}^{A} \right\}  \\
& = \frac{J_{\parallel} + J_{\times}}{2}
\sum_{i} \left( \hat{S}{}_{1,i}^{A} + \hat{\overline{S}}{}_{2,i}^{A}  \right) 
\left(  \hat{S}{}_{1,i+1}^{A} + \hat{\overline{S}}{}_{2,i+1}^{A}  \right)  \\
& \phantom{=}  
+ \frac{J_{\parallel}- J_{\times}}{2} 
\sum_{i}\left( \hat{S}{}_{1,i}^{A} - \hat{\overline{S}}{}_{2,i}^{A}  \right) 
\left(  \hat{S}{}_{1,i+1}^{A} - \hat{\overline{S}}{}_{2,i+1}^{A}  \right)     \; . 
\end{split}
\end{equation}
Combining \eqref{eqn:matrix-elem-rung} and the above, we see that,  
in the strong-coupling limit ($J_{\perp} \to - \infty$), the low-energy effective Hamiltonian is given by: 
\begin{equation}
\mathcal{H}_{\bolN \text{-}\overline{\bolN}}^{\text{eff}} = 
\frac{J_{\parallel} +J_{\times}}{2} \sum_{i} \mathcal{S}^A_{i} \mathcal{S}^A_{i+1} 
+ \frac{J_{\parallel} - J_{\times}}{2} \sum_{i}  \mathcal{D}^A_{i} \mathcal{D}^A_{i+1}  \; .
\label{eqn:strong-coupling-N-Nbar}
\end{equation}
Here we would like to stress that the strong-coupling limit of the usual $\bolN$-$\overline{\bolN}$ two-leg 
SU($N$) ($N \geq 3$) spin ladder \eqref{eqn:def-2leg-ladder} with $J_{\times}=0$ is {\em not} the standard Heisenberg model for the adjoint spins; 
only when $J_{\parallel}=J_{\times}$, we recover the pure Heisenberg model which we know, at least for $N=3$, is in the chiral SPT 
phase.\cite{Morimoto-U-M-F-14}    
As we show in Appendix \ref{sec:DD-int}, 
the SU($N$)-invariant $\mathcal{D}\mathcal{D}$ interaction can be expressed by an eighth-order (sixth-order for $N=3$) 
polynomial in $\{\mathcal{S}^A\}$ and the deviation from $J_{\parallel}=J_{\times}$ brings additional higher-order interactions 
(biquadratic, bicubic, etc.) into the effective Hamiltonian.   For sufficiently large $N$, 
$ \mathcal{D}^A_{i} \mathcal{D}^A_{i+1} \approx (2/N) (\mathcal{S}^A_{i} \mathcal{S}^A_{i+1} )^{2}$ 
[see Eq.~\eqref{eqn:DD-int-by-polynom-gen-N}] and the phase structure in the strong-coupling region can be 
well captured by the following bilinear-biquadratic Hamiltonian:
\begin{equation}
\mathcal{H}_{\text{BLBQ}}^{\text{eff}} = 
\frac{J_{\parallel} +J_{\times}}{2} \sum_{i} \mathcal{S}^A_{i} \mathcal{S}^A_{i+1} 
+ \frac{J_{\parallel} - J_{\times}}{N} \sum_{i}  (\mathcal{S}^A_{i} \mathcal{S}^A_{i+1})^{2}  \; .
\label{eqn:strong-coupling-BLBQr}
\end{equation}
 According to the known results for the SU($N$) bilinear-biquadratic chain in the adjoint representation, 
 the two-fold-degenerate singlet dimer phase (which should not be confused with the non-degenerate RD phase mentioned above) 
dominates when the coefficient of the biquadratic interaction takes sufficiently large negative 
 values.\cite{Morimoto-U-M-F-14,Ueda-M-M-18}  
 This implies that when $J_{\times}$ is sufficiently larger than $J_{\parallel}$ the original ladder system is quadrumerized 
 [as in the crossed dimer (CD) states shown in Fig.~\ref{fig:SD-CD}(a)].  

This is in stark contrast to the case of $N=2$ where both the usual two-leg ladder without the diagonal interaction\cite{Dagotto-R-96} ($J_{\times}$) and 
the composite-spin model ($J_{\times}=J_{\parallel}$) \cite{Solyom-T-86,Schulz-86,Xian-95} lead to the same strong-coupling limit 
except for the overall numerical factor.   
As has been mentioned in Sec.~\ref{sec:intro}, on the line $J_{\parallel}=J_{\times}$, the system may be viewed as 
a single Heisenberg chain in the adjoint representation that is cut into finite segments by singlet rungs ($J_{\perp}$ plays  
the role of ``chemical potential'' for these singlets); when $J_{\perp}<0$ is large enough (strong-coupling limit), the singlet rungs are 
squeezed out and the uniform Heisenberg chain [i.e., the model \eqref{eqn:strong-coupling-N-Nbar} with 
$J_{\times}=J_{\parallel}$] is obtained.    
It is known numerically that the spin chain \eqref{eqn:strong-coupling-N-Nbar} 
on the $J_{\times}=J_{\parallel}$ line is in the chiral SPT phases at least for $N=3$\cite{Morimoto-U-M-F-14} and 
$N=4$\cite{Ueda-M-M-18}.  
Therefore, for sufficiently large negative $J_{\perp}$, the original ladder model \eqref{eqn:def-2leg-ladder} with $J_{\times}=J_{\parallel}$ 
exhibits the chiral SPT order.  
In analogy to a similar problem in the $S=1/2$ ladder,\cite{Xian-95} 
we expect a first-order transition driven by the energetics to occur between the RD (large positive $J_\perp$) 
and the chiral SPT phases (large negative $J_\perp$) 
on the line $J_{\times}=J_{\parallel}$.   In Sec.~\ref{sec:large-N}, we shall show that, in the large-$N$ limit, 
the chiral SPT phase for large negative $J_{\perp}$ is replaced by a period-2 (i.e., plaquette, in the ladder language) singlet phase 
as we increase $J_{\times}$ (see Fig.~\ref{fig:large-N}).  
This implies that the $\mathcal{D}\mathcal{D}$ interaction in the strong-coupling effective Hamiltonian \eqref{eqn:strong-coupling-N-Nbar}   
drives the SPT-dimer (SPT-CD, in the ladder language) transition.   
\section{Large-$N$ limit}
\label{sec:large-N}
To carry out large-$N$ expansion\cite{Affleck-SUN-PRL-85,Read-S-NP-89}, 
it is convenient to express the original Hamiltonian \eqref{eqn:def-2leg-ladder} as:
\begin{equation}
\begin{split}
& \frac{2}{N} \mathcal{H}_{\bolN \text{-}\overline{\bolN}} =  \\
& - J_{\times}  \sum_{j} 
\left\{
P^{(\bolN,\overline{\bolN})}_{\text{singlet}} (1,j;2,j+1)   
+ P^{(\bolN,\overline{\bolN})}_{\text{singlet}} (2,j;1,j+1)  
\right\}  \\
& - J_{\perp} 
\sum_{j} P^{(\bolN,\overline{\bolN})}_{\text{singlet}} (1,j;2,j) \\
& + \frac{1}{N} J_{\parallel} \sum_{j} 
\left\{
\mathcal{P} (1,j;1,j+1)  + \overline{\mathcal{P}} (2,j;2,j+1) 
\right\}   + \text{const.}  
\; 
\end{split} 
\label{eqn:def-N-Nbar-ladder-2}
\end{equation}
where $P^{(\bolN,\overline{\bolN})}_{\text{singlet}}(a,i;b,j) $ ($a,b=1,2$) denotes the projection operator 
onto SU($N$)-singlet on a pair of spins $(a,i)$ and $(b,j)$,  
and $\mathcal{P} (a,i;b,j)$ [$\overline{\mathcal{P}} (a,i;b,j)$] permutes  
the spin states in $\bolN$ ($\overline{\bolN}$) on the sites $(a,i)$ and $(b,j)$ (see Appendix \ref{sec:peojection-permutation}).  
It is important to note that the interactions ($J_{\parallel}$) connecting spins in the identical representation ($\boldsymbol{N}$-$\boldsymbol{N}$ 
or $\overline{\boldsymbol{N}}$-$\overline{\boldsymbol{N}}$) acquire an extra factor $1/N$ and may be treated as 
perturbations in the $1/N$-expansion.  
\subsection{Leading order}
\label{sec:large-N-leading}
\begin{figure}[ht]
\begin{center}
\includegraphics[width=\columnwidth,clip]{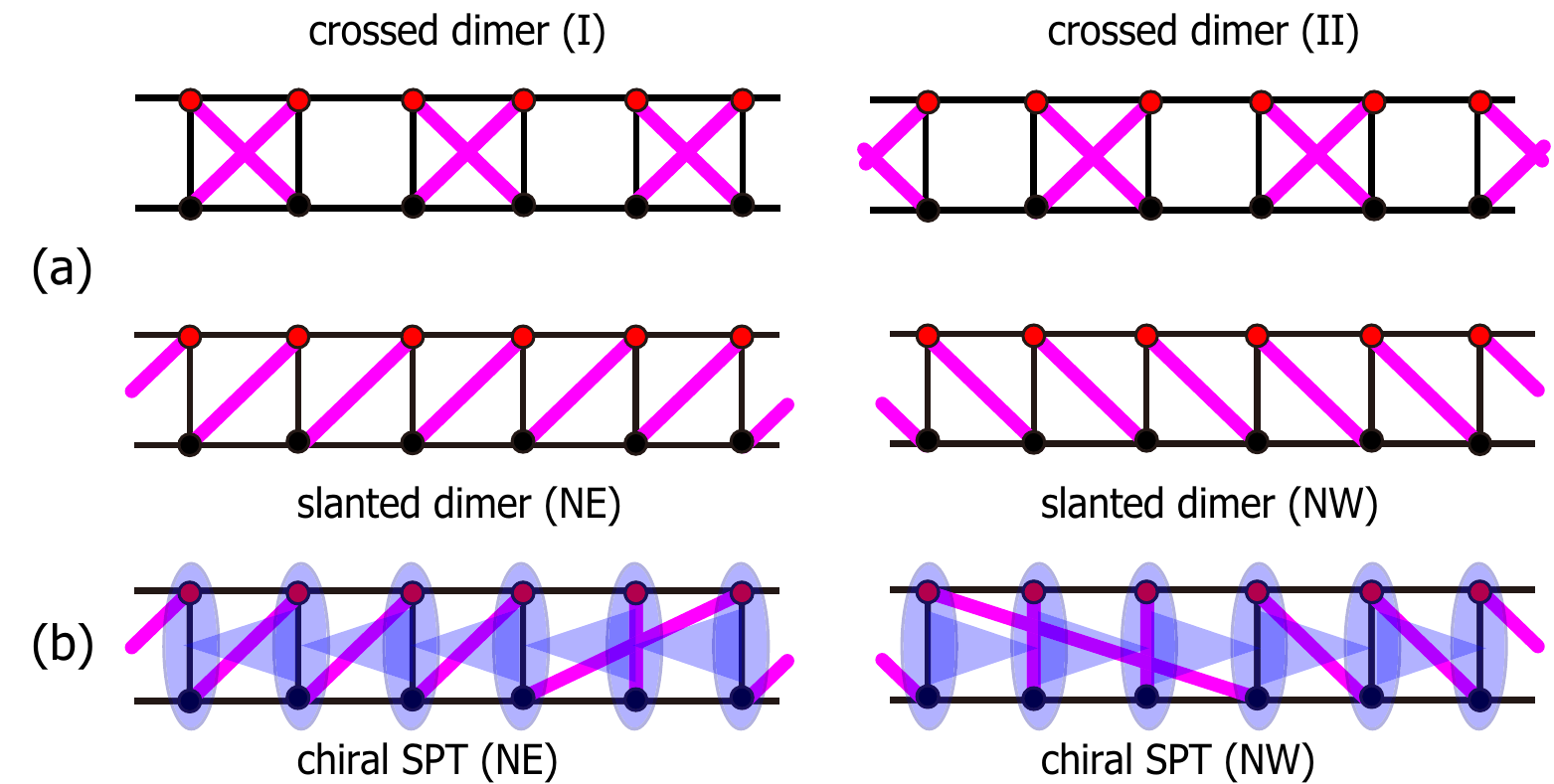}
\caption{
(a) Four degenerate states at the leading order in $1/N$ when $J_{\times} > J_{\perp}$.
Crossed dimer (CD) and slanted dimer (SD) phases. In CD, translation symmetry is broken spontaneously, while 
in SD, reflection symmetry is broken. 
(b) Two chiral SPT states constructed out of ``preformed'' adjoint spins on individual rungs.  In terms of valence-bond covering, 
these states consist of many dimer-pairing patterns including long-range ones.}  
\label{fig:SD-CD}
\end{center}
\end{figure}

The leading [$O(1)$] order in $1/N$, the Hamiltonian \eqref{eqn:def-N-Nbar-ladder-2} 
just counts the number of $\boldsymbol{N}$-$\overline{\boldsymbol{N}}$ bonds in the singlet state:
\begin{equation}
E_{\text{g.s.}} = - J_{\times} \mathcal{N}_{\text{d}} - J_{\perp} \mathcal{N}_{\text{v}}  
\end{equation}  
with $\mathcal{N}_{\text{d}}$ ($\mathcal{N}_{\text{v}}$) being the number of singlets on diagonal (vertical) bonds 
in Fig.~\ref{fig:two-ladders}(a).  
As there are two sites on both edges of each singlet and any one site is never  be shared by more than one singlets, 
the numbers of singlets $\mathcal{N}_{\text{d,v}}$ and the lattice sites $\mathcal{N}_{\text{site}}$ 
must obey the following constraint:
\begin{equation}
\mathcal{N}_{\text{site}}  =  2 \mathcal{N}_{\text{d}} + 2 \mathcal{N}_{\text{v}} \; .
\end{equation}
Therefore, the ground states can be found by minimizing
\begin{equation}
E^{(0)}_{\text{g.s.}} = (J_{\perp} - J_{\times}) \mathcal{N}_{\text{d}} - \mathcal{N}_{\text{site}} J_{\perp} /2  \; .
\end{equation}
The answer depends on $(J_{\times},J_{\perp})$ ($J_{\parallel}, J_{\times}>0$ is assumed).  
If $J_{\times} < J_{\perp}$, $E_{\text{g.s.}}$ is minimized when $\mathcal{N}_{\text{d}}=0$ 
and the RD is the only (non-degenerate) ground state with
\begin{equation}
E^{(0)}_{\text{g.s.}} / \mathcal{N}_{\text{site}}  = - J_{\perp}/2  \equiv e_{\text{RD}}^{(0)}  \; .
\end{equation}
If $J_{\times} = J_{\perp}$, on the other hand, 
{\em any} nearest-neighbor singlet (dimer) coverings on the lattice shown in Fig.~\ref{fig:two-ladders}(b) form 
a highly-degenerate ground state manifold. 
The degeneracy is lifted by $O(1/N)$ processes \cite{Read-S-NP-89} (see Sec.~\ref{sec:QDM}) which can be summarized into 
a quantum dimer Hamiltonian \cite{Rokhsar-K-88} without potential terms. 
When $J_{\times} > J_{\perp}$, the choice $\mathcal{N}_{\text{d}} =\mathcal{N}_{\text{site}}/2$ ($\mathcal{N}_{\text{v}}=0$) 
minimizes the energy:
\begin{equation}
E^{(0)}_{\text{g.s.}} / \mathcal{N}_{\text{site}} = (J_{\perp} - J_{\times}) / 2 - J_{\perp} /2  =   - J_{\times} /2  
\equiv e_{\text{CD/SD}}^{(0)} 
\end{equation}
and the four states ``crossed dimer (CD)'' (two-fold degenerate) and ``slanted dimer (SD)'' (two-fold degenerate) 
are degenerate (any other dimer coverings are higher-lying unlike when $J_{\times} = J_{\perp}$), see Fig.~\ref{fig:SD-CD}.  
There is no first-order [i.e., $O(1/N)$] process that connects these four states and we need $O(1/N^2)$ contributions to resolve the degeneracy. 

\subsection{First order in $1/N$}
\label{sec:QDM}
As we have seen above, there is no first-order correction for $J_{\times} \neq J_{\perp}$.  
When $J_{\times}=J_{\perp}$, any nearest-neighbor dimer coverings on the lattice shown in Fig.~\ref{fig:two-ladders}(b) 
form the manifold of degenerate ground states and 
the degeneracy is lifted at $O(1/N)$ by the processes within the ground-state manifold (first-order degenerate perturbation).
The $O(1/N)$ part of the Hamiltonian is nothing but the quantum dimer model (QDM) \cite{Rokhsar-K-88,Read-S-NP-89}:
\begin{equation}
\mathcal{H}_{\text{QDM}} = 
-t \sum_{\square} \left\{
\left| \raisebox{-1.0ex}{\includegraphics[scale=0.16]{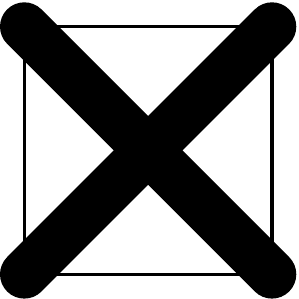}} \right\rangle
\left\langle \raisebox{-1.0ex}{\includegraphics[scale=0.16]{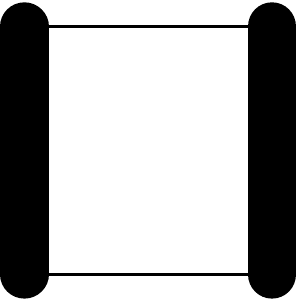}} \right| 
+ \text{h.c.}
\right\} 
\;\; \left( t = \frac{1}{N}(J_{\times} - J_{\parallel} )  \right)   \; , 
\label{eqn:kinetic-QDM}
\end{equation}
where the two resonating plaquette-singlet states are defined as:
\begin{subequations}
\begin{align}
\begin{split}
& \left| \raisebox{-1.0ex}{\includegraphics[scale=0.16]{./figures/N-Nbar-plaquette-CD}}  \right\rangle   \\
& =  \frac{1}{N} \left\{ \sum_{\alpha=1}^{N} | \alpha \rangle_{1,i}\otimes |\bar{\alpha}\rangle_{2,i+1}\right\} \otimes 
\left\{  \sum_{\beta=1}^{N} |\bar{\beta}\rangle_{2,i} \otimes | \beta \rangle_{1,i+1} \right\} 
\end{split}
\\
\begin{split}
& \left| \raisebox{-1.0ex}{\includegraphics[scale=0.16]{./figures/QDM-moves-v}}  \right\rangle   \\
& =  \frac{1}{N} \left\{ \sum_{\alpha=1}^{N}| \alpha \rangle_{1,i}\otimes |\bar{\alpha}\rangle_{2,i}\right\} \otimes 
\left\{  \sum_{\beta=1}^{N} | \beta \rangle_{1,i+1} \otimes |\bar{\beta}\rangle_{2,i+1} \right\}   \; .
\end{split}
\end{align}
\end{subequations}
If we interchange the upper and lower sites on every other rung [see Fig.~\ref{fig:two-ladders}(b)], 
$| \raisebox{-0.5ex}{\includegraphics[scale=0.12]{./figures/N-Nbar-plaquette-CD}} \rangle \to 
 | \raisebox{-0.5ex}{\includegraphics[scale=0.12]{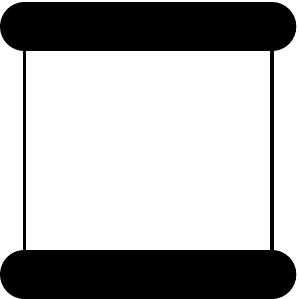}} \rangle$ and 
 Eq.~\eqref{eqn:kinetic-QDM} reduces to the usual kinetic term of the quantum dimer model. For such a purely kinetic term, the ground-state is known to be in a trivial non-degenerate RD phase~\cite{Chepiga-M-19}. 
 The point $J_{\times}=J_{\parallel}(=J_{\perp})$ is very special [the SU($N$)-generalization of the ``composite-spin model''] in that 
the rung-singlet state is the exact eigenstate of the full ladder Hamiltonian \eqref{eqn:def-2leg-ladder}.   
Correspondingly, the above first-order correction vanishes identically.  

\subsection{Second-order correction}
\begin{figure}[ht]
\begin{center}
\includegraphics[width=\columnwidth,clip]{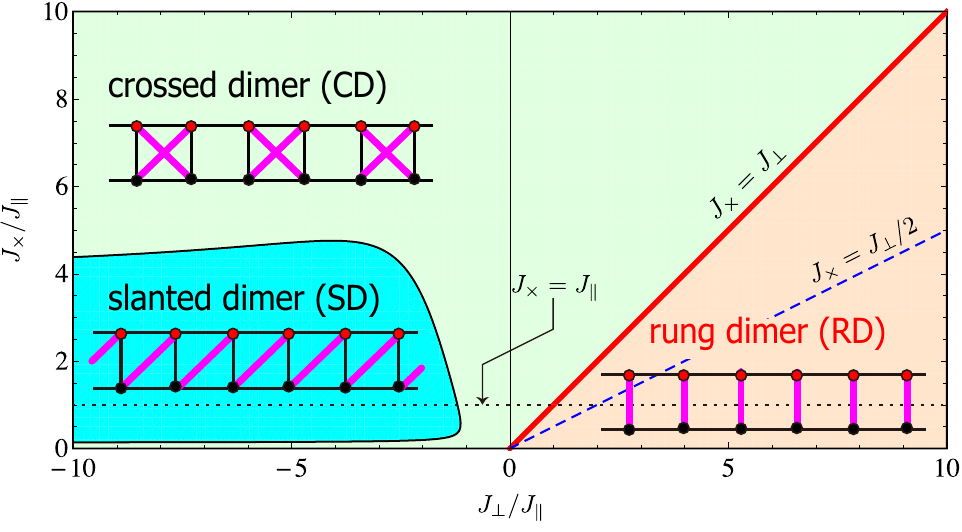}
\caption{
Phase diagram of $\bolN$-$\overline{\bolN}$ ladder \eqref{eqn:def-2leg-ladder} (with $J_{\parallel} >0$) in the large-$N$ limit. 
On the line $J_{\times}=J_{\perp}$, an effective quantum dimer model \eqref{eqn:kinetic-QDM} emerges at $O(1/N)$. 
On the equivalent lattice shown in Fig.~\ref{fig:two-ladders}(b), CD and SD respectively translate to columnar dimer and staggered dimer. 
The two degenerate SD states can be interpreted as two chiral SPT states with non-trivial edge states.  }
\label{fig:large-N}
\end{center}
\end{figure}

As we have seen in Sec.~\ref{sec:large-N-leading}, when $J_{\times} > J_{\perp}$, 
there are four degenerate ground states at the leading order in the large-$N$ expansion and we need to proceed to 
the second order in $1/N$ to resolve the degeneracy.  
Away from the line $J_{\times}=J_{\perp}$, the energies up to $O(1/N^2)$ are 
given as:
\begin{subequations}
\begin{align}
\begin{split}
& E_{\text{CD}}/\mathcal{N}_{\text{site}}  =  - J_{\times}/2  \\
& + \frac{1}{2 N^2} \left\{ 
- 2 J_{\times} 
+ J_{\parallel} 
-\frac{J_{\parallel}^2}{2 J_{\times}} 
+\frac{(J_{\parallel}-J_{\perp}) (J_{\parallel}-J_{\times})}{J_{\perp}-J_{\times}}  \right\} 
\end{split}
\\ 
\begin{split}
& E_{\text{SD}}/\mathcal{N}_{\text{site}}  \\
& = 
- J_{\times}/2  
+ \frac{1}{2 N^2} \left\{ 
- 2 J_{\times} 
- 4 J_{\parallel} 
+ \frac{2 J_{\times} (J_{\perp}-6 J_{\parallel})}{J_{\perp} -2 J_{\times}}  
\right\}  
\end{split}
\\
\begin{split}
& E_{\text{RD}}/\mathcal{N}_{\text{site}}  \\
& = 
- J_{\perp}/2 
-\frac{1}{2 N^2} 
\frac{2 (J_{\parallel} - J_{\perp}) (J_{\parallel} - J_{\times})}{J_{\perp}-J_{\times}}  \; .
\end{split}
\end{align}
\end{subequations} 
The resulting phase diagram is shown in Fig.~\ref{fig:large-N}.    

To relate the large-$N$ results to the phases at finite-$N$, 
we express the SU($N$)-singlet columnar dimer state 
\begin{equation}
\begin{split}
& \left| \raisebox{-1.0ex}{\includegraphics[scale=0.16]{./figures/N-Nbar-plaquette-CD}}  \right\rangle   \\
& =  \left\{ \frac{1}{\sqrt{N}} \sum_{\alpha}| \alpha \rangle_{1a}\otimes |\bar{\alpha}\rangle_{2b}\right\} \otimes 
\left\{ \frac{1}{\sqrt{N}} \sum_{\beta} |\bar{\beta}\rangle_{1b} \otimes | \beta \rangle_{2a} \right\} 
\end{split}
\end{equation}
as a sum of the rung-singlet and the singlet made of two adjoints.  
Using Eq.~\eqref{eqn:adj-states-by-3-3bar}, it is easy to see that 
the (normalized) singlet state made of a pair of preformed adjoint spins (shown by the thick blue lines) is written as:
\begin{equation}
\frac{1}{\sqrt{N^2-1}} \sum_{A=1}^{N^2-1} |A\rangle_{1} \otimes |A\rangle_{2} 
\equiv  \left| \raisebox{-1.0ex}{\includegraphics[scale=0.16]{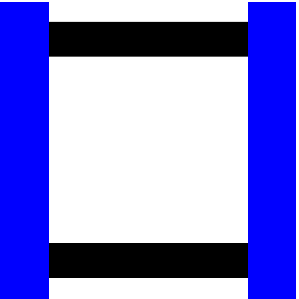}}  \right\rangle  \; .
\end{equation}
Then, by the explicit expression \eqref{eqn:adj-states-by-3-3bar} of $|A\rangle$ 
and the identity 
\begin{equation}
\sum_{A=1}^{N^{2}-1} (S^A)_{\alpha\beta} (S^A)_{\mu \nu}  =  \frac{1}{2} \left(
\delta_{\alpha \nu} \delta_{\beta \mu} - \frac{1}{N} 
\delta_{\alpha \beta} \delta_{\mu \nu} \right)  \; , 
\label{eqn:GaGa-tensor}
\end{equation}
we can express the CD state as:
\begin{equation}
\left| \raisebox{-1.0ex}{\includegraphics[scale=0.18]{./figures/N-Nbar-plaquette-CD}}  \right\rangle
= \frac{\sqrt{N^2-1}}{N}  \left| \raisebox{-1.0ex}{\includegraphics[scale=0.18]{./figures/N-Nbar-plaquette-adj-sing}}  \right\rangle
+ \frac{1}{N} \left| \raisebox{-1.0ex}{\includegraphics[scale=0.18]{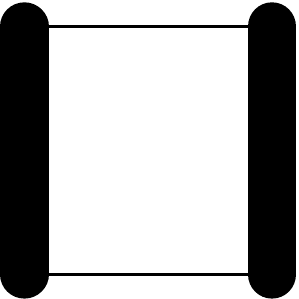}}  \right\rangle 
\xrightarrow{N \to \infty} \left| \raisebox{-1.0ex}{\includegraphics[scale=0.18]{./figures/N-Nbar-plaquette-adj-sing}}  \right\rangle
\end{equation}
which implies that, in the large-$N$ limit, the CD state is dominated by the dimerized state of the (effective) adjoint spins 
on the individual rungs.   

Similar relations hold for the two classes of entangled states (SD and chiral SPT), which are conveniently represented 
with matrix-product states (MPS) as $\otimes_{i} \mathcal{A}^{\text{SD-$\ell$}} (i)$ [for SD; 
see \eqref{eqn:MPS-SD-1} and \eqref{eqn:MPS-SD-2}] and 
$\otimes_{i} \mathcal{M}^{\text{AKLT}}$ [for chiral SPT using a generalization of the Affleck-Kennedy-Lieb-Tasaki (AKLT) construction~\cite{Affleck-K-L-T-88}; see \eqref{eqn:AKLT-by-SD}].  
From Eq.~\eqref{eqn:AKLT-by-SD}, we see that the two SD states are related to the two (reflection-broken) chiral SPT  
states as:
\begin{equation}
\begin{split}
& \mathcal{A}^{\text{SD-NE}} (i)_{ab} \\
& = \frac{\sqrt{N^2 -1 }}{N} \mathcal{M}^{\text{AKLT-NE}}(i)_{ab} 
+ \frac{1}{N} \delta_{ab} \left( \frac{1}{\sqrt{N}} \sum_{c=1}^{N} |c \rangle_{1}|\bar{c} \rangle_{2} \right)  \\
& \mathcal{A}^{\text{SD-NW}} (i)_{ab} \\
& = \frac{\sqrt{N^2 -1}}{N}\mathcal{M}^{\text{AKLT-NW}}(i)_{ab} 
+ \frac{1}{N} \delta_{ab} \left( \frac{1}{\sqrt{N}} \sum_{c=1}^{N} |c \rangle_{1}|\bar{c}\rangle_{2} \right) 
\end{split}
\end{equation}
implying the SD states connect adiabatically to the chiral SPT states in the large-$N$ limit. 
Therefore, the transition between the SD phase and the CD phase 
in the two-leg ladder in the large-$N$ limit may be viewed as an SPT-dimer phase transition occurring 
in the effective Heisenberg chain \eqref{eqn:strong-coupling-N-Nbar} by varying the strength 
of the $\mathcal{D}^A_{i} \mathcal{D}^A_{i+1} $-interaction.

\section{Low-energy description of chiral SPT phase}
\label{sec:weakcoupling}
In this section, we consider the model \eqref{eqn:def-2leg-ladder} for weak interchain interactions and 
show how we describe the stabilization of the SPT phase with broken parity symmetry 
(the chiral SPT phase) within the framework of low-energy effective field-theories.
\subsection{Continuum limit}
\label{sec:continuum-limit}
Without interchain coupling ($J_{\perp},J_{\times} = 0$), the model (\ref{eqn:def-2leg-ladder}) 
becomes two decoupled SU($N$) Sutherland models.  As has been mentioned already in the introduction, 
the latter is integrable and displays a quantum critical behavior in the SU($N$)$_1$ universality
class with the central charge $c=N-1$. \cite{Sutherland-75,Affleck-88}  
The low-energy properties of the Sutherland model for the spins in the $\boldsymbol{N}$ 
(respectively $\overline{\boldsymbol{N}}$) representation can be obtained by starting from
the U($N$) Hubbard chain \eqref{hubbardSUN} at $1/N$ [respectively  $(N-1)/N$] filling and then taking the limit of large repulsive 
interaction $U$. \cite{Affleck-NP86,Affleck-88,James-K-L-R-T-18,Itoi-K-97,Assaraf-A-C-L-99,Manmana-H-C-F-R-11}
At the energy scale well below the charge gap, the SU($N$) operators on the legs $l=1$ and $2$ in the continuum limit are
expressed as: \cite{Affleck-NP86,Affleck-88,James-K-L-R-T-18,Itoi-K-97,Assaraf-A-C-L-99}
\begin{equation}
\begin{split}
S^{A}_{1,j}/a_0 \simeq& J^{A}_{1, \text{L}} (x) +  J^{A}_{1,\text{R}} (x) \\ 
& + \mbox{e}^{ i 2k_{1\text{F}} x} N^{A}_1 (x)  + \mbox{e}^{-i 2k_{1\text{F}} x} N^{A \dagger}_1 (x)   \\
\overline{S}^{A}_{2,j}/ a_0 \simeq& J^{A}_{2, \text{L}} (x) +  J^{A}_{2,\text{R}} (x)   \\
&+  \mbox{e}^{ i 2k_{2\text{F}} x} N^{A}_2 (x) + \mbox{e}^{-i 2k_{2\text{F}} x} N^{A \dagger}_2 (x) ,
\end{split}
\label{spinop}
\end{equation}
where $x = j a_0$ and $k_{1\text{F}} = \pi/(Na_0)$ ($a_0$ being the lattice spacing)
and  $k_{2\text{F}} = \pi (N-1)/(Na_0)$ (see Appendix \ref{Appbosonization} for the details).
In Eq.~\eqref{spinop}, $J^{A}_{l, \text{L,R}}$ denote the left and right SU($N$)$_1$ currents,\cite{Knizhnik-Z-84} 
and the $2k_{lF}$ part involves the SU($N$)$_1$ Wess-Zumino-Novikov-Witten (WZNW) primary field $g_l$ (in $\boldsymbol{N}$) with scaling dimension $(N-1)/N$ (see Appendix \ref{Appbosonization}): 
\begin{eqnarray}
N^{A}_l =  i  \mathcal{C}   \; {\rm Tr} ( g_l  S^A),
 \label{2kf}
\end{eqnarray}
where $\mathcal{C}$ is a non-universal real constant that results from the average 
over the gapped charge degrees of freedom (see Appendix \ref{Appbosonization} for
more details), and $S^A$ are the SU($N$) generators in the fundamental representation defined in Sec.~\ref{sec:strongcoupling}.  
The one-site translation symmetry $\mathcal{T}_{a_0}$ ($S^{A}_{1,i} \rightarrow S^{A}_{1,i+1}$ or $\overline{S}^{A}_{2,i} \rightarrow \overline{S}^{A}_{2,i+1}$) 
acts differently on the $g_{1,2}$ fields due to the difference in the filling factors of the two chains:
 \begin{equation}
\mathcal{T}_{a_0}: \; \; g_1 \rightarrow e^{i 2 \pi  /N} g_1 ,  \; \; g_2 \rightarrow e^{-i 2 \pi  /N} g_2 .
\label{trans}
\end{equation}
The effect of the site-parity symmetry $\mathcal{P}_{\text{s}}$ (or the inversion symmetry such that 
$S^{A}_{1,i} \rightarrow S^{A}_{1,-i}$ and $\overline{S}^{A}_{2,i} \rightarrow \overline{S}^{A}_{2,-i}$) on the $g_{1,2}$ fields can be deduced from the low-energy expressions 
\eqref{spinop} and \eqref{2kf}:
\begin{equation}
\begin{split}
& N^{A}_l (x) \xrightarrow{\mathcal{P}_{\text{s}}}  N^{A \dagger}_l (-x)  \\
& g_l (x) \xrightarrow{\mathcal{P}_{\text{s}}}  - g^{\dagger}_l (-x)  \; .
\end{split}
 \label{siteparity}
\end{equation}

With these basic facts in hands, one can derive the continuum limit of the two-leg spin ladder (\ref{eqn:def-2leg-ladder}) 
when the chains are weakly coupled $|J_{\perp},J_{\times}| \ll J_{\parallel}$.
The leading part of the continuum Hamiltonian reads as
\begin{equation}
\begin{split}
{\cal H} =& \mathcal {H}^{\text{SU($N$)}_1}_{1} + \mathcal{H}^{\text{SU($N$)}_1}_{2}   \\
&+ \lambda_1  \int dx \; \left[ \text{Tr}(g_1g_2) + \text{H.c.} \right]   \\
& + \lambda_2 \int dx \;  \left[ \text{Tr}\, g_1\text{Tr}\, g_2 + \text{H.c.} \right] ,  
\end{split}
 \label{cft}
\end{equation}
where ${\cal H}^{\text{SU($N$)}_1}_{l=1,2}$ denote the Hamiltonians of the SU($N$)$_1$ WZNW CFT for  on the chains $l=1,2$,  
and we have identified the two coupling constants as: 
 \begin{equation}
\lambda_1 = - \left\{J_{\perp}  + 2 J_{\times} \cos(2 \pi/N) \right\} \mathcal{C}^2 a_0  /2, \; \; \lambda_2 = - \lambda_1/N .  
 \label{identcouplings}
\end{equation}
The model \eqref{cft} describes two SU($N$)$_1$ WZNW models perturbed by two strongly relevant perturbations
with the same scaling dimensions $x=2(N-1)/N (< 2)$, which open the spin gap $|\lambda_i |^{N/2}$ (up to logarithmic 
corrections) as soon as the inter-chain interactions are switched on (the gap opens more slowly for larger $N$).  
In Eq. (\ref{cft}), we have not included the marginal current-current interaction:
\begin{equation}
V_{\text{cc}} = \frac{1}{2} \left( J_{\perp} + J_{\times}  - \gamma \right) I^A_{\text{R}} I^A_{\text{L}}
 - \frac{1}{2} \left(J_{\perp} + J_{\times} +\gamma \right) K^A_{\text{R}} K^A_{\text{L}},
\label{Vcc}
\end{equation}
where  the sum $I^A_{\text{R,L}} = J^A_{1,\text{R,L}} + J^A_{2,\text{R,L}}$ is the SU($N$)$_2$ currents 
and the difference $K^A_{\text{R,L}} = J^A_{1, \text{R,L}} - J^A_{2, \text{R,L}}$  is the so-called wrong currents.\cite{Gogolin-N-T-book}
In Eq.~\eqref{Vcc}, the coupling $\gamma > 0$ comes from the marginally irrelevant in-chain current-current perturbation,  
which is responsible for the logarithmic corrections to the SU($N$)$_1$ criticality of the individual Sutherland models. \cite{Itoi-K-97} 

The effective Hamiltonian \eqref{cft} and the marginal interactions \eqref{Vcc} thus describe the low-energy physics of the two-leg spin 
ladder \eqref{eqn:def-2leg-ladder}  in the weak-coupling regime $|J_{\perp},J_{\times}| \ll J_{\parallel}$.   It takes a rather different form from the one obtained in Ref.~\onlinecite{Lecheminant-T-15} [see Eq.~(9) there] 
where the case of the two chains in the identical representation $\boldsymbol{N}$ is considered, \footnote{%
In the case of identical chains, we have $g_2^{\dagger}$ (i.e., primary field for $\overline{\boldsymbol{N}}$) instead of $g_2$ (for $\boldsymbol{N}$).}  though the perturbing fields with the same scaling dimensions appear in both cases.  
From the expressions \eqref{identcouplings} of the coupling constants, 
we observe that the leading interactions in the low-energy effective Hamiltonian \eqref{cft} disappear 
if we fine-tune $J_{\perp} =-2 J_{\times} \cos(2 \pi/N)$ ($J_{\perp} =0$ for $N=4$).   
Then, on top of the marginal current-current interactions \eqref{Vcc}, strongly irrelevant perturbation with non-zero conformal spin 
of the generic form $N^{A}_1 \partial^2_x N^{A \dagger}_2 + \text{H.c.}$ must be taken into account.    
Unfortunately, as is well known,\cite{Gogolin-N-T-book} no decisive conclusion can be drawn for the latter kind of perturbation.   
The relevant perturbations of Eq.~\eqref{cftnewbasis} will be generated in higher-orders of perturbation theory 
as in the $N=2$ case\cite{Starykh-B-04,Kim-L-S-08,Hikihara-S-10} since the field theory \eqref{cft} 
with independent $\lambda_{1,2}$ couplings is the most general one allowed by SU($N$) invariance, site parity,  
and translational invariance.  
Although we will not precisely investigate the physics along these special lines in the following, 
we naively expect that these special lines describe the vicinity of the quantum phase transitions between competing orders 
as the numerical phase diagrams suggest.

\subsection{Conformal-embedding approach}
We now try to elucidate the physical nature of the phases in the weak-coupling limit  by exploiting the existence of the following conformal embedding\cite{Bais-B-S-S-88}  
as in Ref.~\onlinecite{Lecheminant-T-15}:
\begin{equation}
\text{SU($N$)}_1 \times \text{SU($N$)}_1 \sim \text{SU($N$)}_2 \times  \mathbb{Z}_N ,
\label{embedding}
\end{equation}
where ${\mathbb{Z}}_N$ is the parafermionic CFT with central charge $c=2(N-1)/(N+2)$ which
describes the universal properties of the phase transition of the two-dimensional $\mathbb{Z}_N$
Potts model. \cite{Zamolodchikov-F-JETP-85}
The SU($N$)$_2$ CFT has the central charge $c=2(N^2-1)/(N+2)$ and is generated by the currents  $I^A_{ R,L}$.
The two  SU($N$)$_1$  WZNW fields $g_{1,2 }$ can be expressed in the new 
SU($N$)$_2$ $\times$ ${\mathbb Z}_N$ basis as:\cite{Griffin-N-89,Lecheminant-T-15}
\begin{equation}
\begin{split}
& (g_1)_{\alpha \beta} \sim  G_{\alpha \beta} \; \sigma_1   \\
& (g_2)_{\alpha \beta} \sim  G_{\alpha \beta}  \;  \sigma_1^{\dagger} , 
\end{split}
\label{ident}
\end{equation}
where $\alpha, \beta = 1, \ldots, N$ and $G$ is the SU($N$)$_2$  WZNW field (in $\boldsymbol{N}$-representation)  
with scaling dimension $\Delta_G = (N^2-1)/N(N+2)$. In Eq. (\ref{ident}), $ \sigma_1$ is the ${\mathbb Z}_N$ spin field 
with scaling dimension $\Delta_{\sigma} = (N-1)/N(N+2)$. \cite{DiFrancesco-M-S-book} From Eqs. \eqref{trans} and 
\eqref{siteparity}, we can find the continuous description of the one-site translation symmetry $\mathcal{T}_{a_0}$ 
and the site-parity transformation in the new SU($N$)$_2 \times  \mathbb{Z}_N$  basis.
Using the identification \eqref{trans}, we observe that the one-site translation symmetry $\mathcal{T}_{a_0}$ 
does not act on the SU($N$)$_2$ WZNW $G$ field but brings about the ${\mathbb Z}_N$ transformation on the spin field: 
\begin{equation}
\sigma_1 \xrightarrow{\mathcal{T}_{a_0}} e^{2 \pi i/N} \sigma_1 \; .  
\label{eqn:1-site-tr-in-Zn}
\end{equation}
On the other hand, the site-parity transformation \eqref{siteparity} coincides with the conjugation for $G$ and $\sigma_1$ fields: 
\begin{equation}
\begin{split}
& G (x) \xrightarrow{\mathcal{P}_{\text{s}}}  G^{\dagger} (-x)  \\
&  \sigma_1 (x) \xrightarrow{\mathcal{P}_{\text{s}}}  - \sigma_1^{\dagger}  (-x) \; .
\end{split}
 \label{siteparityembedding}
\end{equation}

With all these results, we can express the non-interacting and interacting parts of the model \eqref{cft} 
in the new SU($N$)$_2 \times  \mathbb{Z}_N$ basis respectively as:
\begin{equation}
\begin{split}
& \mathcal {H}^{\text{SU($N$)}_1}_{1} + \mathcal{H}^{\text{SU($N$)}_1}_{2}  
= \mathcal {H}^{\text{SU($N$)}_2} + \mathcal {H}^{{\mathbb Z}_N}  \;  ,  \\
& \mathcal{H}_{\text{int}} =  {\tilde \lambda}_1  \int dx \left\{  \left(\text{Tr} \, G\right)^2 +  \text{Tr} \, G^2 + \text{H.c.} \right\}  \\
& \phantom{\mathcal{H}_{\text{int}} = }  
+ {\tilde \lambda}_2 \int dx    \left\{ \left(\text{Tr} \, G\right)^2  -  \text{Tr} \, G^2  + \text{H.c.} \right\}  \epsilon_1 \; , 
\end{split}
\label{cftnewbasis}
\end{equation}
with the following set of coupling constants: 
\begin{equation}
\begin{split}
& {\tilde \lambda}_1 = -  \frac{N - 1}{2N} \left\{ J_{\perp} + 2 J_{\times} \cos(2 \pi/N) \right\} \mathcal{C}^2  a_0  \\
& {\tilde \lambda}_2 =   \frac{N + 1}{2N}  \left\{J_{\perp} + 2 J_{\times} \cos(2 \pi/N)\right\} \mathcal{C}^2  a_0   \; .
\end{split}
\label{lambda12diag}
\end{equation}
The marginal current-current contribution \eqref{Vcc} can also be expressed in the new basis: 
\begin{equation}
V_{\text{cc}} = \frac{1}{2} \left(J_{\perp} + J_{\times}  - \gamma \right) I^A_{ R} I^A_{ L}
 + \frac{1}{2} \left(J_{\perp} + J_{\times} +\gamma \right)  \epsilon_1 \text{Tr} \, \Phi_{\rm adj} \; , 
\label{Vccnewbasis}
\end{equation}
where $\Phi_{\rm adj}$ is the SU($N$)$_2$ primary field in the adjoint representation with scaling dimension $2N/(N+2)$ and $ \epsilon_1$ is the first thermal operator, singlet under the ${\mathbb Z}_N$ symmetry, with scaling
dimension $\Delta_{\epsilon} = 4/(N+2)$. \cite{DiFrancesco-M-S-book}

\subsection{Field-theory description of chiral SPT phase}
\label{sec:chiral-SPT-by-FT}
The resulting effective field theory is different from the one  for the two-leg SU($N$) spin ladder made of two identical 
chains\cite{Lecheminant-T-15} in that none of the two strongly relevant perturbations in Eq.~\eqref{cftnewbasis} 
is integrable [see Eq.~(27) in Ref.~\onlinecite{Lecheminant-T-15}, where the perturbation in the ${\mathbb Z}_N$ sector 
is integrable].   
Therefore, one expects, on general grounds, that these terms will produce spin-gapped phases 
with different physical properties from the one found in Ref.~\onlinecite{Lecheminant-T-15}.
To shed light on the possible fully gapped phases of the original lattice model (\ref{eqn:def-2leg-ladder}), 
we first single out the relevant SU($N$) perturbation in Eq.~\eqref{cftnewbasis} with coupling constant $ {\tilde \lambda}_{1}$ 
which does not involve the ${\mathbb Z}_N$ degrees of freedom. A spectral gap $\Delta_{\text{s}}$ is formed in the SU($N$) sector and, after averaging over the SU($N$) sector, we obtain an effective theory for the SU($N$)-singlet ${\mathbb Z}_N$ degrees of freedom that governs the physics at the energy scale $E \ll \Delta_{\text{s}}$.  
This phenomenological approach is expected to capture the spin-gapped phases of the weakly-coupled two-leg SU($N$) spin ladder (\ref{eqn:def-2leg-ladder}) in the regime where the spin-gap $\Delta_{\text{s}}$ is much larger than that 
in the singlet ${\mathbb Z}_N$ sector.  

We focus our analysis on the description of the chiral SPT or the SD phase in Fig.~\ref{fig:large-N} 
found  within the large-$N$ limit by considering the situation $ {\tilde \lambda}_{1} > 0$ [i.e. $J_{\perp} < - 2 J_{\times} \cos(2 \pi/N)$].  The strongly-relevant perturbation $\text{Tr} \, G^2 +  \left(\text{Tr} \, G\right)^2+ \text{H.c.}$ of Eq. (\ref{cftnewbasis}) opens a gap $\Delta_{\text{s}}$ and its minimization when $ {\tilde \lambda}_{1} > 0$ gives  
a variety of solutions depending on $N$ ($N > 2$):
\begin{equation}
\begin{split}
G= 
\begin{cases}
\pm i I  & (N=4p \geq 4)   \\
e^{\pm i 2 p \pi/N} I  &  (N=4p+1 \geq 5)  \\
 \pm i \; {\rm diag} (1,1,1,1,1,-1) &  (N=6)   \\
\pm i \; e^{\pm i  \pi /N}   I   &  (N=4p+2 \ge 10)   \\
e^{\pm i (N+1) \pi/2N} I  &   (N=4p+3 \geq 3)  ,
\end{cases}
\end{split}
\label{MinPhiS}
\end{equation}
with $I$ being the $N$-dimensional identity matrix.  
It is important to note that all these solutions break the conjugation symmetry $G \to G^{\dagger}$ for all $N$ 
and thus the site-parity ($\mathcal{P}_{\text{s}}$) symmetry (\ref{siteparityembedding}).  
It thus opens the way to the description of a chiral SPT phase within low-energy field-theory approaches.

Averaging over the $G$ field in Eqs.~\eqref{cftnewbasis} and \eqref{Vccnewbasis}, 
we obtain the following effective action for the SU($N$)-singlet ${\mathbb Z}_N$ parafermion sector:
\begin{equation}
{\cal S}_{\rm eff} = {\cal S}_{{\mathbb Z}_N} +  {\tilde \lambda} \int d^2 x  \; \; \epsilon_1 ,
\label{effactionparaJperpF}
\end{equation}
where $ {\cal S}_{{\mathbb Z}_N}$ is the Euclidean action of the ${\mathbb Z}_N$  CFT\cite{Zamolodchikov-F-JETP-85}   
with the coupling constant defined by:
\begin{equation}
\begin{split}
& {\tilde \lambda} =  \frac{N + 1}{2N}  \left\{ J_{\perp} + 2 J_{\times} \cos(2 \pi/N) \right\} \mathcal{C}^2 a_0 \\
&\times \left\langle \left( \left(\text{Tr} \, G\right)^2  -  \text{Tr} \, G^2  + \text{H.c.} \right) \right\rangle 
+  \frac{a_0}{2} \left(J_{\perp} + J_{\times} +\gamma \right) \langle  \mbox{Tr} \Phi_{\rm adj} \rangle \\
&= - \left\{ J_{\perp} + 2 J_{\times} \cos(2 \pi/N) \right\}  A +  \left(J_{\perp} + J_{\times} +\gamma \right) B  
\end{split}
\label{lambdatildebis}
\end{equation}
(the bracket $\langle \cdots \rangle$ denotes the average over the gapped SU($N$) sector
and the two constants $A$ and $B$ are positive).  
The  ${\mathbb Z}_N$ effective action \eqref{effactionparaJperpF} is integrable
and describes a massive field theory for all sign of ${\tilde \lambda}$. \cite{Fateev-91}  When $ {\tilde \lambda} >0$, we have $\langle  \epsilon_1 \rangle <0$ and, in our convention, the underlying two-dimensional ${\mathbb Z}_N$ lattice model belongs 
to its high-temperature (paramagnetic) phase where the  ${\mathbb Z}_N$ symmetry is restored.  
In our two-leg spin ladder, it means that the one-step translation symmetry $\mathcal{T}_{a_0}$ \eqref{eqn:1-site-tr-in-Zn} 
is not spontaneously broken.  
On the other hand, the site-parity symmetry $\mathcal{P}_{\text{s}}$ \eqref{siteparityembedding} 
is spontaneously broken [see Eq.~\eqref{MinPhiS}] and two-fold degenerate ground states result. 
In this respect, the resulting spin-gap phase when ${\tilde \lambda} >0$ may be identified with the chiral SPT or the SD phase 
discussed in Sec.~\ref{sec:large-N} within the large-$N$ limit approach. This exotic phase emerges in the phase diagram of generalized two-leg spin ladder (\ref{eqn:def-2leg-ladder}) within the weak-coupling regime when
\begin{equation}
\begin{split}
& J_{\perp} < - 2 J_{\times} \cos(2 \pi/N) \; ,  \\
&  \left(J_{\perp} + J_{\times} +\gamma \right) B  > \left\{ J_{\perp} + 2 J_{\times} \cos(2 \pi/N) \right\}  A  .
\end{split}
\label{constraintSPTchiralweak}
\end{equation}
Even within the weak-coupling region, mapping out the full phase diagram needs the precise knowledge of 
the relative values of the non-universal positive parameters $A$, $B$  and $\gamma$ in Eq.~\eqref{constraintSPTchiralweak} 
which makes the actual analysis rather difficult.  In this respect, a direct numerical investigation that will be carried out 
in the next section is indispensable to the precise determination of the phases.  Nevertheless, in some special cases,  
one can conclude the emergence of the chiral SPT phase without knowing the explicit values of the parameters $A$, $B$  and $\gamma$.   
For instance, when $N=3$, one observes from \eqref{constraintSPTchiralweak} that 
$J_{\times} > J_{\perp}$ and $J_{\times} > - J_{\perp}$ are sufficient for ${\tilde \lambda}_1>0$ and ${\tilde \lambda} >0$ to hold.   
Our low-energy approach predicts then the formation of the chiral SPT phase {\em at least} in this region of couplings. 
A similar conclusion can be drawn for $N=4$ when $J_{\perp} < 0$ and $J_{\times} > - J_{\perp} $.  
These are to be compared with the phase diagrams Figs.~\ref{fig:dmrg_su3} and \ref{fig:dmrg_su4} obtained numerically 
in the next section.  

A more precise connection can be established between 
the chiral SPT phase described within the low-energy approach and the SD phase found in the large-$N$ limit  
by considering, for instance, the special case $N=3$ with $J_{\perp} =0$ and  $J_{\times} > 0$ 
(in which the above analysis predicts the chiral SPT phase). 
To this end, we add the following small anisotropy 
\begin{equation}
\mathcal{V} =  K_{\text{diag}}  \sum_i \left( S^{A}_{1,i}  \overline{S}^{A}_{2,i+1}   -  \overline{S}^{A}_{2,i} S^{A}_{1,i+1}  \right) 
\label{2leganisodiag}
\end{equation}
between the spin-exchange along the two diagonals of the original ladder 
(\ref{eqn:def-2leg-ladder}) which explicitly breaks the site-parity $\mathcal{P}_{\text{s}}$ 
thereby selecting one of the two degenerate chiral SPT states.  
At low energies, using Eq. \eqref{spinop} and the identification \eqref{ident}, 
we can derive the continuum-limit expression of $\mathcal{V}$ as:
\begin{equation}
\begin{split}
\mathcal{V} \simeq&  \kappa_1  \int dx  \; i\left\{ \left(\text{Tr} \, G\right)^2 +   \text{Tr} \, G^2 - \text{H.c.} \right\}   \\
&+  \kappa_2 \int dx    \; i \left\{ \left( \text{Tr} \, G\right)^2  -  \text{Tr}\, G^2  - \text{H.c.} \right\}  \epsilon_1  ,
\end{split}
\label{Vnewbasis}
\end{equation}
with the two coupling constants given by:
\begin{equation}
\kappa_1 =  \frac{\sqrt{3} K_{\text{diag}} }{6} \mathcal{C}^2 a_0 \; , \;\; 
\kappa_2 = - \frac{\sqrt{3} K_{\text{diag}}  }{3} \mathcal{C}^2 a_0    \; .
\label{kappa12diag}
\end{equation}
When $K_{\text{diag}} >0$, one has $\kappa_1 > 0 $ and the minimization of the first term in Eq.~\eqref{Vnewbasis} 
uniquely selects one of the two semiclassical solutions in Eq.~\eqref{MinPhiS} $G = e^{-i 2\pi/3} I$ that breaks $\mathcal{P}_{\text{s}}$ 
while keeping the ${\mathbb Z}_3$ symmetry unbroken ($\langle  \epsilon_1 \rangle <0$).
When  $K_{\text{diag}} <0$, on the other hand, we get the site-parity partner $G = e^{i 2\pi/3}  I$.  
In either case, broken $\mathcal{P}_{\text{s}}$ and unbroken $\mathcal{T}_{a_0}$ 
indicate that the solution found in each case corresponds to one of the chiral SPT phases 
and may be identified, for sufficiently large $|K_{\text{diag}}| \gg 1$, with the SD phase shown in the lower panels of Fig.~\ref{fig:SD-CD}(a) 
that is characterized by the formation of SU(3) singlets along one of the diagonals of the ladder.   
When open-boundary condition is used, the chiral SPT phase with the edge states (${\bf \bar 3}, {\bf 3}$) 
[respectively (${\bf 3},{\bf \bar 3}$)] emerges for $K_{\text{diag}}>0$ (respectively $K_{\text{diag}}<0$).

\section{DMRG calculations}
\label{sec:dmrg}
In this section, we map out the phase diagram of the generalized two-leg spin ladder (\ref{eqn:def-2leg-ladder}) for 
$N=3$ and $N=4$ by means of  DMRG calculations.\footnote{We have used the ITensor C++ library,  
available at \url{http://itensor.org} for our calculations.}  
Using the $(N-1)$ U(1) conserved quantities (corresponding to the $N-1$ Cartan generators which we call ``colors'')    
and keeping up to $m=4000$ states, we were able to obtain converging results for 
ladders of length $L=240$ for $N=3$ and $L=128$ for $N=4$ with a truncation error below $10^{-6}$ in the worst cases.  
Note that we will fix $J_\parallel=1$ as the unit of energy. 

For later convenience, we briefly describe how we label the topological phases protected by on-site {\em projective} SU($N$) 
[i.e., $\text{PSU($N$)}$] symmetry.\cite{Duivenvoorden-Q-13}  If the state is invariant under 
on-site symmetry $G=\text{PSU($N$)}$, 
the corresponding SPT phases are labeled by the projective representations $\mathcal{R}_{\text{e}}$ of $G$  
under which the edge states transform.   
Since inequivalent projective representations of PSU($N$) are labeled by the number of the boxes $n_{\text{Y}}$ (mod $N$) 
appearing in the Young diagram of $\mathcal{R}_{\text{e}}$\cite{Duivenvoorden-Q-13},  
there are $(N-1)$ non-trivial topological classes (as well as one trivial one) specified by the $\mathbb{Z}_{N}$ label 
$n_{\text{top}}=n_{\text{Y}}$ (mod $N$).   In PSU($N$) ($N \geq 3$), there is an ambiguity in labeling the topological class by the edge states.  
If the right edge states transform like $\mathcal{R}_{\text{e}}$ in a given topological phase, the left ones like its conjugate 
$\overline{\mathcal{R}}_{\text{e}}$ which is distinct from $\mathcal{R}_{\text{e}}$, in general.  
In the following, we define the topological class by the {\em right edge state} and use the name ``class-$n_{\text{top}}$''  
for the topological phases with $n_{\text{top}}=n_{\text{Y}}$ (mod $N$)  (with the class-0 corresponding to trivial phases).  
For instance, a pair of chiral SPT phases with ($\overline{\bolN},\bolN$) and  ($\bolN,\overline{\bolN}$) belong 
respectively to class-1 and class-($N-1$), while the reflection-symmetric phase $({\bf 6},{\bf 6})$ 
identified in Refs.~\onlinecite{Nonne-M-C-L-T-13,Bois-C-L-M-T-15,Capponi-L-T-16} for SU(4) is called 
class-2.  

To characterize the phases, we measured all local quantities, i.e. color-resolved site densities 
(i.e., the eigenvalues of the $N-1$ Cartan generators), bond energies on the two legs, rungs, and the two diagonal bonds,  
as well as the entanglement spectrum when cutting the ladder in the middle. In particular, the topological class 
can be easily identified by determining the edge states which transform according 
to some specific SU($N$) representations~\cite{Fromholz-C-L-P-T-19}.

To probe the SPT phase, it is also convenient to measure some bulk correlations 
using the string-order parameters (SOP)~\cite{denNijs-R-89,Duivenvoorden-Q-12,Duivenvoorden-Q-ZnxZn-13} 
which allows direct access to the topological label $n_{\text{top}}$ (i.e., the projective representations 
under which the edge states transform).\cite{Pollmann-T-12,Hasebe-T-13} 
As is discussed in Appendix \ref{sec:string-OP}, the building blocks of them are a pair of $\mathbb{Z}_{N}$ generators 
$(\hat{P},\hat{Q})$ and the operators $(\hat{X}_P,\hat{X}_Q)$ transforming like $\mathbb{Z}_{N}$ order parameters.  
Following Refs.~\onlinecite{Duivenvoorden-Q-ZnxZn-13} and  \onlinecite{Tanimoto-T-15}, we define
\begin{equation}
  \mathcal{O}_1(m,n) = \lim_{|i-j|\rightarrow \infty} \left\langle \{\hat{X}_P(i)\}^m \left\{ \prod_{k=i}^{j-1} \hat{Q}_k^n \right\}  
  \{\hat{X}^\dagger_P(j)\}^m \right\rangle
  \label{eq:sop}
\end{equation}
where the sets of integers $(m,n)$ are chosen in such a way that the set $\{ {\cal O}_1(m,n) \}$ distinguishes among  
all $(N-1)$ possible SPT phases for a given $N$.  
It is known\cite{Duivenvoorden-Q-ZnxZn-13,Tanimoto-T-15} that when the SOP $\mathcal{O}_1(m,n)$ is non-zero, 
$(m,n)$ and the topological class $n_{\text{top}}$ must be related to each other by: 
\begin{equation}
m+n \, n_{\text{top}} = 0 \quad   (\text{mod $N$})  \; .
\label{eqn:selection-rule-SOP}
\end{equation}   
For more details, see Appendix \ref{sec:string-OP}.

Furthermore, to make connection with the effective Hamiltonian \eqref{eqn:strong-coupling-N-Nbar} 
that corresponds to the strong-coupling limit $J_\perp \rightarrow -\infty$ of the original model \eqref{eqn:def-2leg-ladder}, 
we also directly simulated this SU($N$) chain where each site transforms in the adjoint representation. 

\subsection{SU(3) case}
Regarding the SOP in the $N=3$ case, we choose $\mathcal{O}_1(1,2)$ and $\mathcal{O}_1(1,1)$ 
that satisfy \eqref{eqn:selection-rule-SOP} to detect the class-1 and 2 SPT phases, respectively.   
The pair of $\mathbb{Z}_{N}$ generators $Q$ and $P$ for the defining representation $\boldsymbol{N}$ are 
given in Appendix \ref{sec:string-OP}.    
To adapt these results for our $\mathbf{3}$-$\bar{\mathbf{3}}$ ladder, we need 
to define all the operators in \eqref{eq:sop} on the individual rungs and 
use $Q_{\text{rung}}(\mathbf{3} \otimes \bar{\mathbf{3}}) = Q(\mathbf{3}) \otimes Q(\bar{\mathbf{3}})$ 
and $\hat{X}_{P,\text{rung}}(\mathbf{3} \otimes \bar{\mathbf{3}}) = \hat{X}_P(\mathbf{3}) \otimes \mathbf{1} 
+ \mathbf{1} \otimes \hat{X}_P(\bar{\mathbf{3}})$ with 
\begin{equation}
\begin{split}
& Q({\bf 3}) =
 \begin{pmatrix}
1 & 0 & 0 \\
 0 &  \omega & 0 \\
 0 & 0 & \omega^{2} 
 \end{pmatrix}
, \; 
X_P({\bf 3}) =
\frac{i}{\sqrt{3}}
\left(
\begin{array}{c c c}
  \omega^{-1} & 0 & 0 \\
  0 & \omega & 0 \\
  0 & 0 & 1
\end{array}
  \right)
\\
& Q(\bar{\mathbf{3}}) =
\begin{pmatrix}
  1 & 0 & 0 \\
 0 &  \omega^2 & 0 \\
 0 & 0 & \omega
\end{pmatrix}
,
X_P(\bar{\mathbf{3}}) =
\frac{-i}{\sqrt{3}}
\left(
\begin{array}{c c c}
  \omega^{2} & 0 & 0 \\
  0 & \omega & 0 \\
  0 & 0 & 1
\end{array}
\right)
\end{split}
\label{eqn:Q-P-Xp-Xq-SU3}
\end{equation}
and $\omega=\exp(i 2\pi/3)$ (see Appendix \ref{eqn:ZnxZn-gen-by-spins} for the derivation).  
In our DMRG computations, we fixed the rungs $i$ and $j$ in \eqref{eq:sop} at positions $L/4$ and $3L/4$ respectively so that 
we can measure the bulk correlations free from the contributions due to the edge states.
  
\begin{figure}[ht]
\begin{center}
\includegraphics[width=\columnwidth,clip]{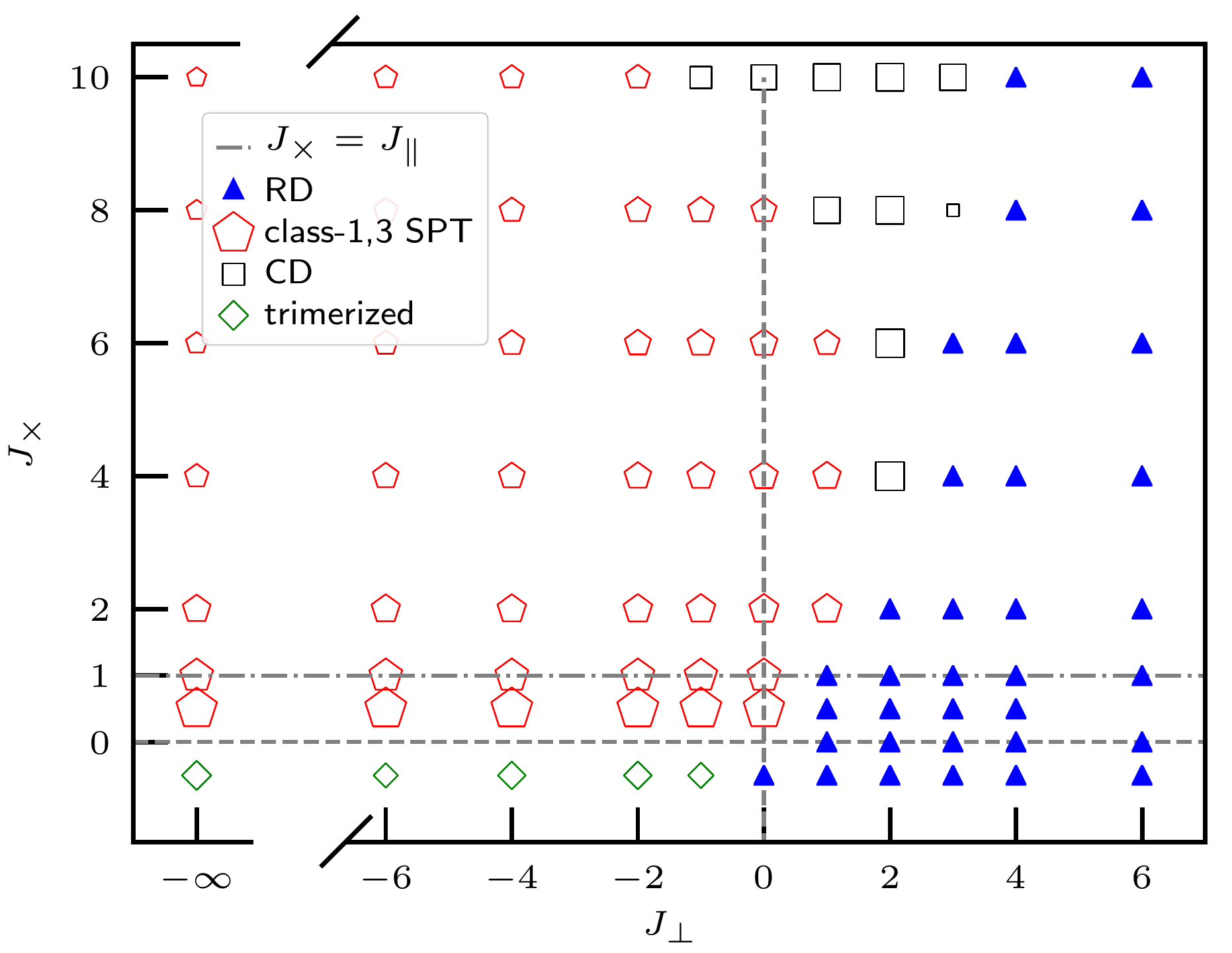}
\caption{
Phase diagram for $N=3$ vs ($J_{\perp},J_{\times}$) obtained from DMRG simulations on ladders of length $L=240$. Except in the trivial RD phase (which has no order parameter), symbols' sizes are proportional to the values of each order parameter, see text. For completeness, data obtained for the strong-coupling effective Hamiltonian \eqref{eqn:strong-coupling-N-Nbar} are also plotted and correspond to the limit $J_\perp \rightarrow -\infty$. 
Note that the fine-tuned line $J_{\perp}=J_{\times}$ predicted by 
field theory seems to describes the SPT-RD boundary in the weak-coupling region.
\label{fig:dmrg_su3}}
\end{center}
\end{figure}

We start by showing the numerical phase diagram in Fig.~\ref{fig:dmrg_su3}, which was obtained by DMRG simulations on ladders of length $L=240$ by measuring various bulk order parameters described below.  
According to the field-theory analysis in Sec.~\ref{sec:weakcoupling}, the origin $(J_{\perp},J_{\times})=(0,0)$ 
is an isolated critical point corresponding to two copies of $\text{SU($3$)}_1$ WZNW CFTs [with $c=4$] 
and the spin gap opens as soon as we deviate from the origin in any generic directions.  
First of all, the large $J_{\perp}(>0)$ region is dominated by the trivial RD phase (plotted by triangles in Fig.~\ref{fig:dmrg_su3}) 
where all the above order parameters vanish.  
This is very different from the phase diagram of the two-leg ladder with identical chains in which we have an extended critical 
phase described by $c=2$ $\text{SU(3)}_1$ WZNW CFT at large enough $J_{\perp}>0$.\cite{Weichselbaum-C-L-T-L-18}  
On top of the trivial RD phase, we found the following gapped phases:\\
\underline{\em CD phase}: following Fig.~\ref{fig:large-N}, we can identify CD phase simply by measuring the difference between diagonal bond strengths in the middle of the ladder
\begin{equation*}
\left\langle S^{A}_{1,L/2}  \overline{S}^{A}_{2,L/2+1} \right\rangle  -  \left\langle S^{A}_{1,L/2+1} \overline{S}^{A}_{2,L/2+2}  \right\rangle  \;  .
\end{equation*} 
\\
\underline{\em SPT phase}:  
in the numerical phase diagram Fig.~\ref{fig:dmrg_su3}, 
we show the values of the SOP, i.e. the maximum of ${\cal O}_1(1,1)$ and ${\cal O}_1(1,2)$, by the size of the hexagons.  
We also checked the reflection-symmetry breaking associated with the chiral SPT phases 
by measuring the difference between two diagonal bonds in a central plaquette
\begin{equation*}
\left\langle S^{A}_{1,L/2}  \overline{S}^{A}_{2,L/2+1} \right\rangle  -  \left\langle \overline{S}^{A}_{2,L/2} S^{A}_{1,L/2+1} \right\rangle  \;  .
\end{equation*}
The SPT phases are stabilized in a large part of the phase diagram.  
In fact, the region is much larger than the one ($J_{\times} > |J_{\perp}|$) predicted by field theory in Sec.~\ref{sec:chiral-SPT-by-FT} 
(this is not surprising as the field-theory prediction has been made as the sufficient condition).   
 \\
\underline{\em Trimerized phase}:   one can test also whether the system spontaneously breaks translation symmetry into a trimerized phase.  
To this end, we measured the corresponding order parameter
  $$ \langle S^{A}_{1,L/2}  S^{A}_{1,L/2+1} \rangle -\frac{1}{2}\langle S^{A}_{1,L/2+1}  S^{A}_{1,L/2+2} \rangle -\frac{1}{2}\langle S^{A}_{1,L/2+2} 
 S^{A}_{1,L/2+3} \rangle $$ 
for trimerization.

\begin{figure}[ht]
\begin{center}
\includegraphics[width=\columnwidth,clip]{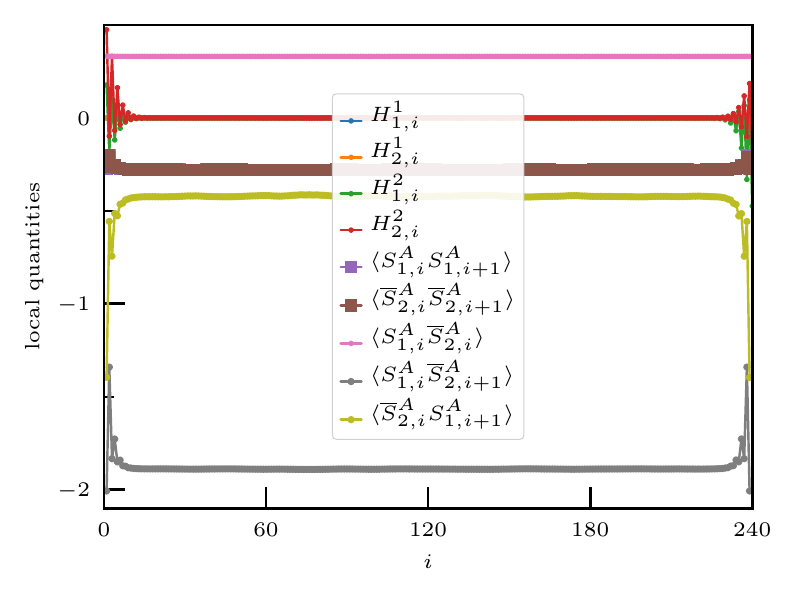}
\caption{Local quantities obtained by DMRG simulations on ladders of length $L=240$ for $J_\perp=-6$ and $J_\times=0.5$.
  We can see a large difference in the diagonal bond strengths, which is a direct evidence of reflection-symmetry 
  breaking in the chiral SPT phase. By summing up the averages of the diagonal (Cartan) generators 
  around the right edge, we can detect the class-1 SPT phase, with edge state weights $(H_1,H_2)=(0.0,-0.816) \simeq (0,-2/\sqrt{6})$ corresponding to one of the three states in the fundamental representation of SU(3). 
\label{fig:dmrg_su3_local}}
\end{center}
\end{figure}

\begin{figure}[ht]
\begin{center}
\includegraphics[width=0.325\columnwidth,clip]{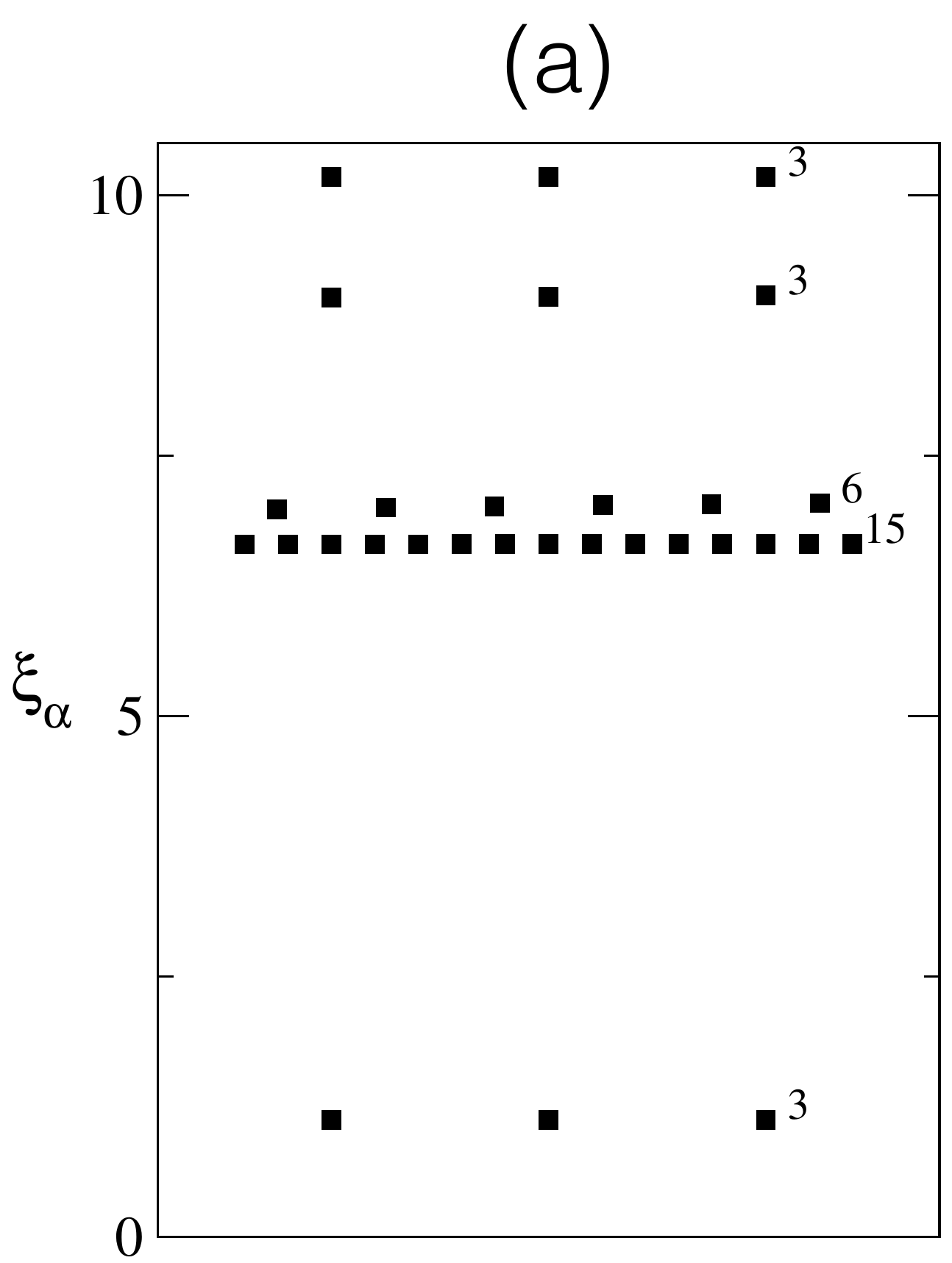}
\includegraphics[width=0.325\columnwidth,clip]{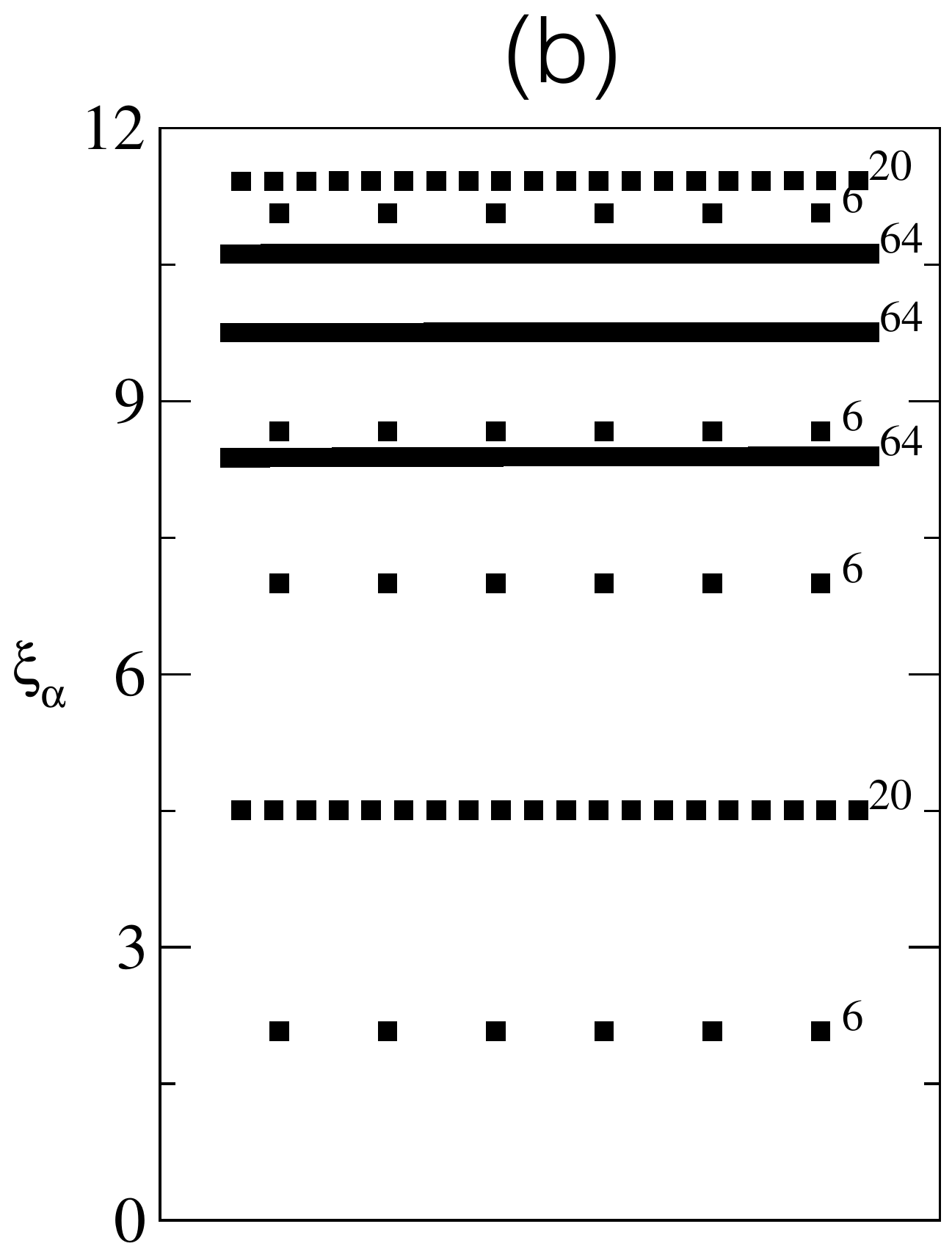}
\includegraphics[width=0.325\columnwidth,clip]{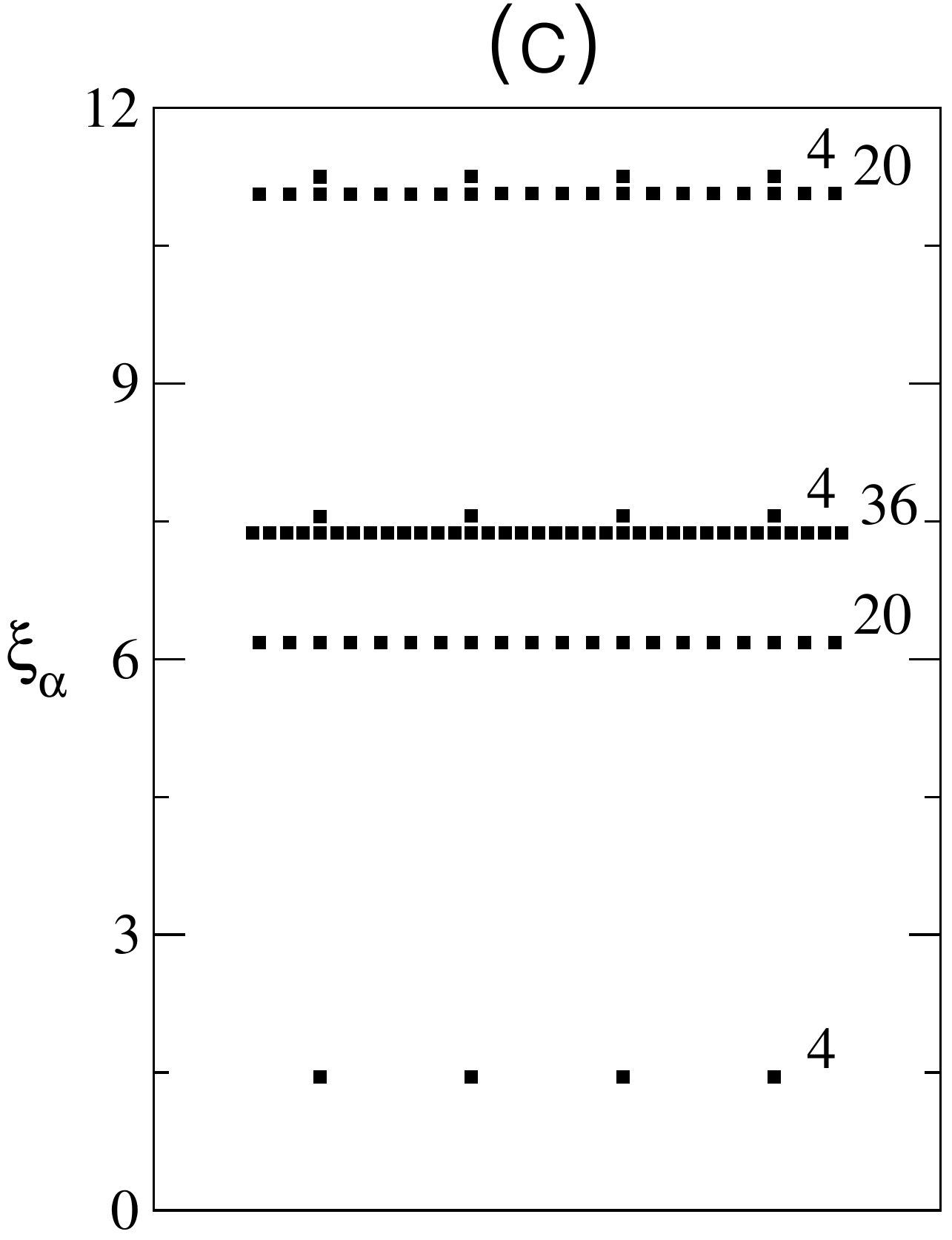}
\caption{Entanglement spectra obtained by DMRG simulations for a half-system. (a) $N=3$ SPT phase with $J_\perp=-6$ and $J_\times=0.5$. (b) $N=4$ class-2 SPT phase with $J_\perp=-4$ and $J_\times=-1$. (c) $N=4$ class-1 SPT phase 
with $J_\perp=-4$ and $J_\times=1$. Note that the entanglement spectra are in full agreement 
with Refs.~\onlinecite{Morimoto-U-M-F-14,Tanimoto-T-15,Fromholz-C-L-P-T-19,Ueda-M-M-18}.}  
\label{fig:es}
\end{center}
\end{figure}

To illustrate what typical local quantities look like in the SPT phase, 
we plot them in Fig.~\ref{fig:dmrg_su3_local}  for $J_\perp=-6$ and $J_\times=0.5$.  
The characteristics of the chiral SPT phase can be seen in the presence of the left and right edge states 
(which transform as $\mathbf{3}$ and $\bar{\mathbf{3}}$, respectively, thus signaling the chiral SPT phase).  
Note that due to its chiral nature, the SPT phase also breaks reflection symmetry  
so that there is a sizeable difference in the diagonal bond amplitudes 
$|\langle S^A_{1,i} \overline{S}^A_{2,i+1} - \overline{S}^A_{2,i} S^A_{1,i+1}\rangle |$ in the bulk (see Fig.~\ref{fig:SD-CD}).    
We also measured the string order parameters in the bulk and found a non-zero value ${\cal O}_1(1,2) \simeq 0.377$ as expected.  
In such an SPT phase, the entanglement spectrum of a half-system 
also exhibits characteristic features reflecting the edge states, as can be seen in Fig.~\ref{fig:es}(a), 
where the degrees of degeneracy are given by the dimensions of the SU(3) representations 
with Young diagrams that contain $n_{\text{Y}}=1$ (modulo 3) boxes~\cite{Morimoto-U-M-F-14,Ueda-M-M-18}.

\subsection{SU(4) case}
\begin{figure}[ht]
\begin{center}
\includegraphics[width=\columnwidth,clip]{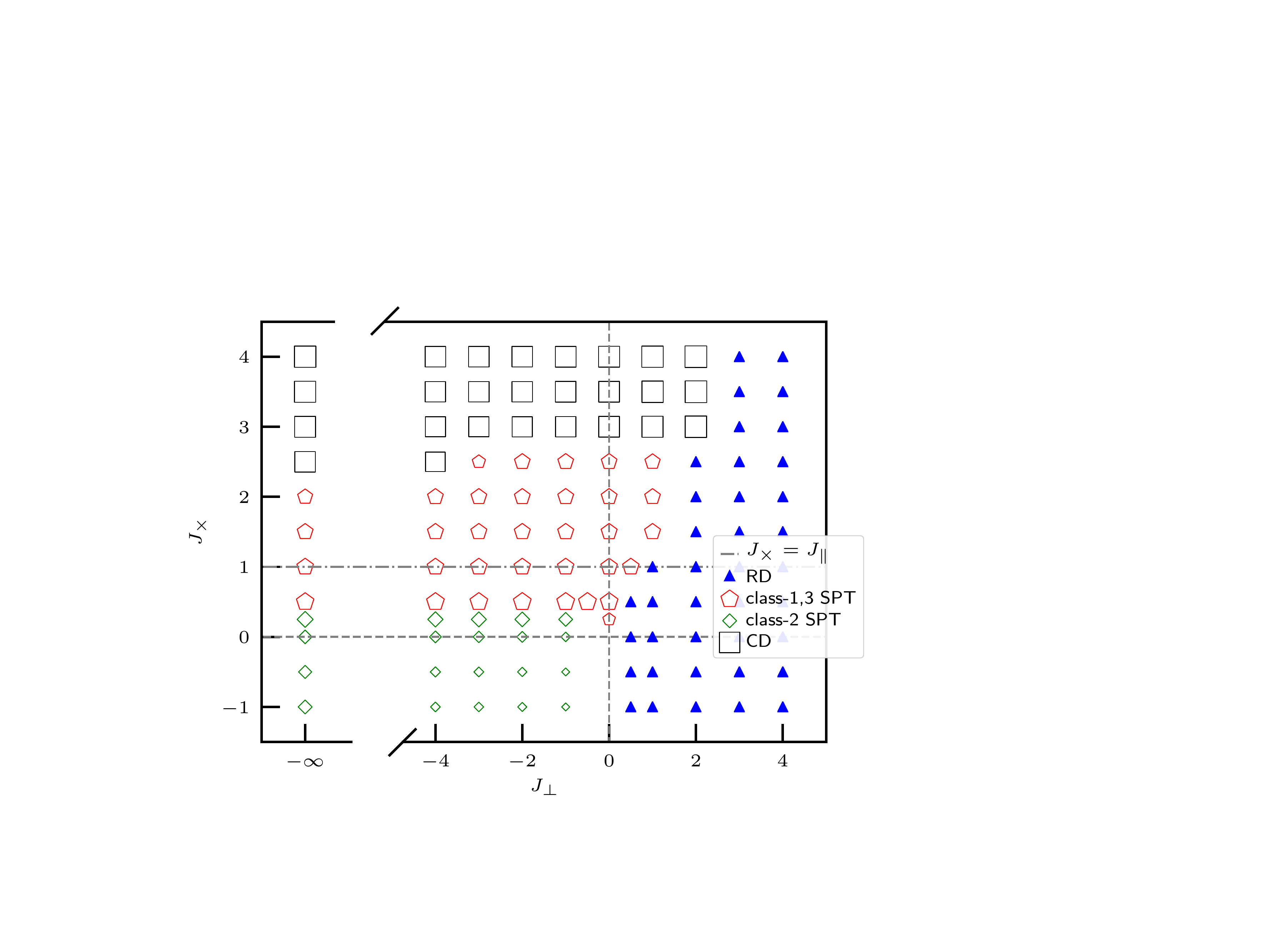}
\caption{
Phase diagram for $N=4$ vs ($J_{\perp},J_{\times}$) obtained from DMRG simulations on ladders of length $L=128$.  
Except in the trivial (RD) phase (which has no order parameter), symbols' sizes are proportional to the values of each order parameter, see text. For completeness, data obtained on the strong coupling effective Hamiltonian \eqref{eqn:strong-coupling-N-Nbar} are also plotted and correspond to the limit $J_\perp \rightarrow -\infty$.
}
\label{fig:dmrg_su4}
\end{center}
\end{figure}

We now turn to the $N=4$ case which is summarized in the phase diagram of Fig.~\ref{fig:dmrg_su4}.
To map out the SPT phases, we computed \eqref{eq:sop} with $(m,n)=(1,3)$, $(2,1)$, and $(1,1)$ 
to detect class-1, 2, and 3 SPT phases, respectively.    
For our $\mathbf{4}$-$\bar{\mathbf{4}}$ ladder, we have the following definitions of the $\mathbb{Z}_{4}$ generator on each rung 
$Q_{\text{rung}}(\mathbf{4} \otimes \bar{\mathbf{4}}) = Q(\mathbf{4}) \otimes Q(\bar{\mathbf{4}})$ and 
the $\mathbb{Z}_{4}$ order parameter 
$\hat{X}_{P,\text{rung}}(\mathbf{4} \otimes \bar{\mathbf{4}}) = \hat{X}_P(\mathbf{4}) \otimes \mathbf{1} + \mathbf{1} \otimes \hat{X}_P(\bar{\mathbf{4}})$ with 
\begin{equation}
\begin{split}
& Q({\bf 4}) =
 \begin{pmatrix}
1 & 0 & 0 & 0\\
 0 &  i & 0 & 0 \\
 0 & 0 & -1 & 0 \\
 0 & 0 & 0 & -i
 \end{pmatrix}
\, , \; 
X_P({\bf 4}) = \frac{1}{2}e^{-i\frac{\pi}{4}} Q^\dagger({\bf 4})
\\
& Q(\bar{\mathbf{4}}) = Q({\bf 4})^*, \quad X_P(\bar{\mathbf{4}}) = -\frac{1}{2}e^{-i\frac{\pi}{4}} Q^\dagger({\bf 4})  \; .
\end{split}
\label{eqn:Q-P-Xp-Xq-SU4}
\end{equation}
As before, in our DMRG computations, we fixed the rungs $i$ and $j$ to positions $L/4$ and $3L/4$, respectively 
so that we can measure the bulk correlations without the edge contributions. 
In the phase diagram shown in Fig.~\ref{fig:dmrg_su4}, the SPT phases were identified by looking at the SOP values 
${\cal O}_1(1,3)$, ${\cal O}_1(2,1)$ and ${\cal O}_1(1,1)$.  The observed numerical values of ${\cal O}_1(1,3)$ or ${\cal O}_1(1,1)$ are significantly larger than those of ${\cal O}_1(2,1)$ (typically a maximal value about $0.2$ of the former compared to 
$0.01$ for the latter). This observation may be naturally understood by the values expected for the model AKLT states 
corresponding to each SPT phase (see Appendix~\ref{sec:string-OP}): using our definitions, the class-1 and 3 AKLT states, 
built from physical adjoint spin at each site, lead to an SOP value of $(8/15)^2 \simeq 0.284$, 
while the class-2 SPT AKLT, which is built with the {\em same} physical spin, exhibits a much smaller value 
$|{\cal O}_1(2,1)|=(2/15)^2 \simeq 0.018$.   

\begin{figure}[ht]
\begin{center}
\includegraphics[width=\columnwidth,clip]{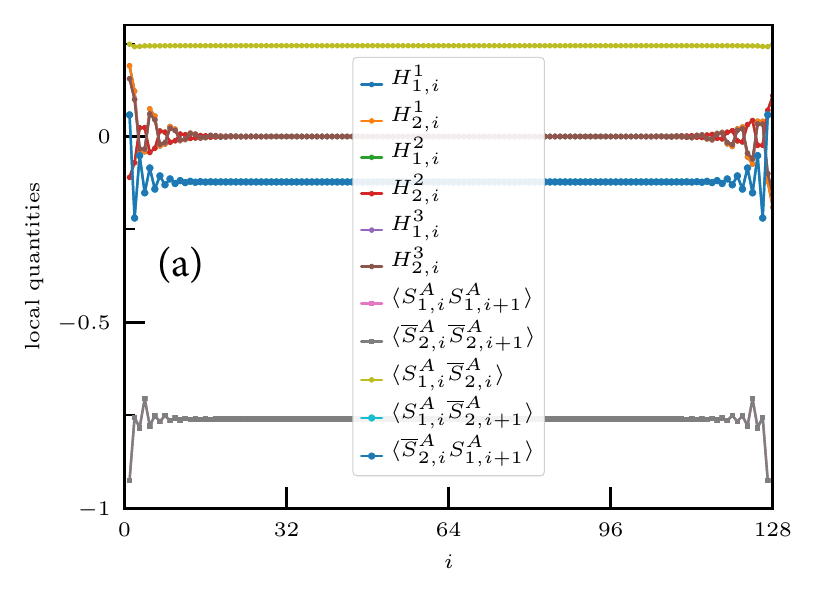}
\includegraphics[width=\columnwidth,clip]{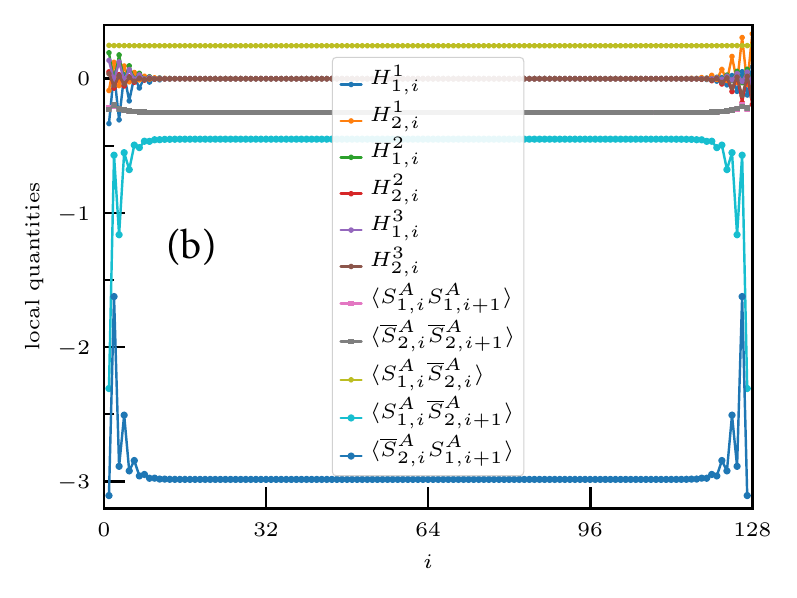}
\caption{
  Local quantities obtained by DMRG simulations on ladders of length $L=128$ for $J_\perp=-4$ and (a) $J_\times=0$ or (b) $J_\times=0.5$.
  By summing up the averages of the (diagonal) Cartan generators over the right edge, we can detect respectively: 
  (a) the class-2 SPT phase, with edge state weights $(H_1,H_2,H_3)=(-0.707,0.408,-0.577)\simeq(-1/\sqrt{2},1/\sqrt{6},-1/\sqrt{3})$ corresponding to one of the states in the self-conjugate antisymmetric ${\bf 6}$ representation of SU(4);
  (b) the class-3 SPT phase, with edge state weights $(H_1,H_2,H_3)=(0.707,-0.408,-0.289)\simeq (1/\sqrt{2},-1/\sqrt{6},-1/\sqrt{12})$ corresponding to one of the 4 states in the ${\bf \bar 4}$ representation of SU(4), 
  and very different diagonal bond amplitudes.}
\label{fig:dmrg_su4_local}
\end{center}
\end{figure}

The phase diagram consists of four major phases (see Fig.~\ref{fig:dmrg_su4}).  
As has been discussed in Sec.~\ref{sec:strongcoupling}, 
for large positive $J_\perp$, the local degrees of freedom on each rung are frozen into an SU(4) singlet.  
As a result, in this part of the phase diagram, there is a large region of a trivial gapped phase dubbed RD in the previous sections 
(see Fig.~\ref{fig:large-N}).   Note that, in the ladder with identical chains, this region is occupied by 
a period-2 valence-bond crystal.\cite{Weichselbaum-C-L-T-L-18}   
For $J_\perp<0$ and $J_\times \sim 0$, on the other hand, we discovered quite surprisingly that 
the class-2 SPT phase with self-conjugate edge states $({\bf 6},{\bf 6})$ is stabilized, which was expected 
neither in the large-$N$ expansion nor in the field theory analysis.   
It turned out that this SPT phase was not stable against a moderate $J_\times \sim 1$  
being replaced by the (${\bf 4},{\bf \bar 4}$) or (${\bf \bar 4}, {\bf 4}$) chiral SPT phases,  
as is expected from the large-$N$ results (Fig.~\ref{fig:large-N}) or the low-energy approach in Sec.~\ref{sec:weakcoupling}.  
This can be seen both in the change in the edge states and in the structure of the entanglement spectrum,  
where the six-fold degeneracy of the lowest level indicative of the class-2 non-chiral $(\mathbf{6},\mathbf{6})$ SPT phase 
[for $J_\perp=-4$ and $J_\times=-1$; Fig.~\ref{fig:es}(b)]
changes to the four-fold degeneracy characteristic of the class-1,3 chiral $(\mathbf{4},\bar{\mathbf{4}})$ SPT phase 
[for $J_\perp=-4$ and $J_\times=1$; Fig.~\ref{fig:es}(c)].  
The transition point $J_{\times}\approx 0.38$ ($J_{\perp} \to -\infty$) 
determined in Ref.~\onlinecite{Ueda-M-M-18} is consistent with ours (data not shown) and moreover does not depend strongly on $J_\perp$ 
(see Fig.~\ref{fig:dmrg_su4}). 
Also since the chiral SPT phases break reflection symmetry spontaneously,  
it can be detected by measuring different diagonal bonds; see Fig.~\ref{fig:dmrg_su4_local}(b).  
This transition appears to be of first order since, even around the transition point, 
the entanglement entropy of a block does not seem to diverge with the size of the block.
Finally, let us emphasize that the chiral SPT phase (class-1 or class-3) is stabilized in a large portion of the phase diagram 
including the region $J_{\perp}<0$ and $J_{\times}  > -J_{\perp}$ predicted by (weak-coupling) field theory in Sec.~\ref{sec:weakcoupling}.  

If one further increases $J_\times$ (for negative or small $J_\perp$), 
a period-2 quadrumerized phase similar to the ``crossed'' dimer (CD) phase found in the large-$N$ limit (Fig.~\ref{fig:SD-CD}), 
where the unit cell contains two rungs, is stabilized.  
As in the SU(3)  case, its order parameter is simply given by the difference in the diagonal bond energies between 
the neighboring plaquettes in the center of the ladder.

\section{Conclusion}
\label{sec:conclusion}
We have studied the zero-temperature phase diagram of a two-leg SU($N$) spin ladder \eqref{eqn:def-2leg-ladder}, 
dubbed $\bolN$-$\overline{\bolN}$ ladder, where the upper and lower legs are made of spins 
in the fundamental ($\boldsymbol{N}$) representation and its conjugate ($\overline{\boldsymbol{N}}$), respectively, 
as a function of parameters $J_\perp$ and $J_\times$ (the leg interaction is set to unity: $J_{\parallel}=1$). 
For odd $N$, such a model can be derived as a low-energy effective Hamiltonian 
of SU($N$) fermions trapped in a 1D array of double wells.~\cite{Fromholz-C-L-P-T-19}

When the inter-chain (or, rung) interaction is much larger than the others $|J_{\perp}| \gg J_{\parallel}, |J_{\times}|$, 
each rung is either in the singlet state (when $J_{\perp}>0$) or in the adjoint (when $J_{\perp}<0$).  
When $J_{\perp}>0$, the system is in the SU($N$)-singlet rung-dimer (RD) phase with gapped $(N^2 -1)$-plet 
excitations that are analogous to the triplon excitations in the usual spin-1/2 ladder.  
When $J_{\perp}<0$, on the other hand, the system is described effectively by a new type of ``spin'' Hamiltonian 
\eqref{eqn:strong-coupling-N-Nbar} in which there is a $\mathcal{D}\mathcal{D}$ interaction as well as the usual 
Heisenberg-like one.  The addition of the $\mathcal{D}\mathcal{D}$ interaction amounts to adding higher-order polynomials  
in the SU($N$) generators.  

To obtain some insights into the global structure of the phases, we have used 
a large-$N$ expansion and shown how conventional phases can emerge, such as trivial rung-dimer (RD) phase, 
or the ``crossed dimer'' (CD) which breaks 1-site translation symmetry.  
On top of these phases, we have found the ``slanted dimer'' (SD) phase which is adiabatically connected to 
the chiral SPT phase with  non-trivial ($\bolN,\overline{\bolN}$) or  ($\overline{\bolN},\bolN$) edge states 
related by the inversion symmetry.  

A low-energy analysis has been carried out in the weak-coupling regime ($|J_{\times}|,|J_{\perp}| \ll |J_{\parallel}|$), 
starting from decoupled SU($N$) chains and then applying the conformal-embedding approach where 
the low-energy effective theory is written in terms of the SU($N$) degrees of freedom and the SU($N$)-singlet 
$\mathbb{Z}_{N}$-parafermion.    
The situation is quite different from that in the usual SU($N$) ladder with an identical ($\boldsymbol{N}$) representation for both legs,~\cite{Lecheminant-T-15} 
and it generically leads to gapped phases, either one with broken translation symmetry 
or topological chiral SPT phases that break reflection symmetry instead.  

A large-scale numerical (DMRG) analysis has fully confirmed these predictions and also provided some quantitative phase diagrams 
for SU(3) (Fig.~\ref{fig:dmrg_su3}) and SU(4) (Fig.~\ref{fig:dmrg_su4}).
For $N=3$, all the approaches consistently point to the conclusion that a simple one-dimensional model \eqref{eqn:def-2leg-ladder} 
hosts all the two non-trivial SPT phases, 
i.e., the (${\bf \bar 3}, {\bf 3}$) chiral SPT phase and its parity-partner (${\bf 3},{\bf \bar 3}$) 
characterized respectively by the topological label $n_{\text{top}}=n_{\text{Y}}=1$ and $2$ (mod $3$) 
in a fairly large portion of the $J_{\perp} \le 0$, $J_{\times} > 0$ region.  

Quite surprisingly, for $N=4$, on top of the class-1 [$(\bar{\mathbf{4}}, \mathbf{4})$] and 3 [$(\mathbf{4}, \bar{\mathbf{4}})$] 
chiral phases that spontaneously break reflection symmetry,  
we have numerically found for small or negative $J_{\times}$ the non-chiral class-2 SPT phase 
with self-conjugate edge states $({\bf 6},{\bf 6})$.   
Therefore, the simple $\bolN$-$\overline{\bolN}$ ladder \eqref{eqn:def-2leg-ladder} with $N=4$ can realize 
all the possible SPT phases, too. 
Besides these chiral and non-chiral SPT phases and the standard RD phase, a fully gapped phases 
with broken translation symmetry [e.g., crossed dimer (CD) phase for all $N$ and the trimerized phase for $N=3$] 
were also found. 

We hope that some of these exotic phases will be explored in future cold atom experiments using ytterbium or strontium atoms for instance.

\begin{acknowledgments}
One of the authors (KT) is supported in part by JSPS KAKENHI Grant No.~15K05211, 18K03455 and No.~JP15H05855.  
This work was performed using HPC resources from GENCI (Grant No. A0050500225) and CALMIP.
Last, the authors thank the program ``Exotic states of matter with SU($N$)-symmetry'' (YITP-T-16-03) held at Yukawa Institute for Theoretical Physics where early stage of this work has been carried out.   
\end{acknowledgments}

\appendix

\section{Some remarks on strong-coupling limit}
\subsection{Strong-coupling limit of two-leg ladder with identical legs}
\label{sec:N-N-ladder-strong-coupling}
To highlight the peculiarity of the $\boldsymbol{N}$-$\overline{\boldsymbol{N}}$ ladder, 
let us consider the following two-leg ladder with {\em identical} spins (i.e., those in the defining representation $\boldsymbol{N}$) 
on both legs:\cite{Lecheminant-T-15,Weichselbaum-C-L-T-L-18} 
\begin{equation}
\begin{split}
\mathcal{H}_{\boldsymbol{N}\text{-}\boldsymbol{N}} = &
J_{\parallel}   \sum_{j} 
\left\{
S_{1,j}^{A} S_{1,j+1}^{A}  + S_{2,j}^{A} S_{2,j+1}^{A} 
\right\}   + 
J_{\perp} 
\sum_{j}\left( 
 S_{1,j}^{A} S_{2,j}^{A} 
\right) \\
&+ J_{\times}  \sum_{j} 
\left\{
S_{1,j}^{A} S_{2,j+1}^{A}  + S_{2,j}^{A} S_{1,j+1}^{A} 
\right\} 
\; .
\end{split}
\end{equation}
As a pair of spins on the rung-$j$ satisfy
\begin{equation}
\sum_{A=1}^{N^2-1} S_{1,j}^{A} S_{2,j}^{A} = 
\begin{cases}
\frac{1}{2N}(N-1)  & \text{symmetric 2-tensor ${\tiny \yng(2)}$} \\
- \frac{1}{2N}(N+1) & \text{anti-symmetric 2-tensor ${\tiny \yng(1,1)}$}  \; , 
\end{cases}
\label{eqn:SdotS-N-N}
\end{equation}
in the strong-coupling limit $|J_{\perp}| \gg J$, the state with symmetric  ${\tiny \yng(2)}$ is chosen for $J_{\perp}<0$ and 
anti-symmetric ${\tiny \yng(1,1)}$ for $J_{\perp}>0$.  
\subsubsection{Case of $J_{\perp}>0$:}
The $N(N-1)/2$ basis states in the anti-symmetric 2-tensor representation may be written as
\begin{equation}
|\alpha,\beta\rangle = \frac{1}{\sqrt{2}} (|\alpha\rangle_{1} \otimes |\beta\rangle_{2} - |\beta\rangle_{1} \otimes |\alpha\rangle_{2}) 
\quad (1 \leq \beta < \alpha \leq N) \; .
\end{equation}
It is straightforward to calculate the following matrix elements:
\begin{equation}
\begin{split}
& \langle \mu,\nu| \hat{S}^{A}_{1}|\alpha,\beta\rangle  \\
&= \frac{1}{2} \left\{ 
(S^{A})_{\mu\alpha}\delta_{\beta\nu} - (S^{A})_{\mu\beta}\delta_{\alpha\nu} 
- (S^{A})_{\nu\alpha}\delta_{\beta\mu} + (S^{A})_{\nu\beta}\delta_{\alpha\mu} 
\right\}  \, ,  \\
& \langle \mu,\nu | \hat{S}^{A}_{2}|\alpha,\beta\rangle
= \langle \mu,\nu| \hat{S}^{A}_{1}|\alpha,\beta\rangle  \; .
\end{split}
\end{equation}
These immediately imply
\begin{equation}
\begin{split}
& \langle \mu,\nu | \hat{S}^{A}({\tiny \yng(1,1)})  |\alpha,\beta\rangle
= \langle \mu,\nu | \left\{ \hat{S}^{A}_{1} + \hat{S}^{A}_{2} \right\} |\alpha,\beta\rangle  \\
& = 2 \langle \mu,\nu | \hat{S}^{A}_{1}|\alpha,\beta\rangle
= 2 \langle \mu,\nu| \hat{S}^{A}_{2}|\alpha,\beta\rangle \; .
\end{split}
\end{equation}
Therefore, within the first-order perturbation, we obtain the following strong-coupling effective (Heisenberg) 
Hamiltonian \cite{vdBossche-A-L-M-01}:
\begin{equation}
\mathcal{H}^{\boldsymbol{N}\text{-}\boldsymbol{N}}_{\text{eff}}(J_{\perp}>0) 
= \frac{1}{2}(J_{\parallel} + J_{\times}) \sum_{j} S^{A}_{j}({\tiny \yng(1,1)}) S^{A}_{j+1}({\tiny \yng(1,1)})  \; , 
\label{eqn:eff-Ham-sym-ladder-AF}
\end{equation}
where $\{ S^{A}_{j}({\tiny \yng(1,1)})\}$ denote the spin operators in the representation ${\tiny \yng(1,1)}$.    

\subsubsection{Case of $J_{\perp}<0$:}
Now the single-rung ground state is in the symmetric 2-tensor representation ${\tiny \yng(2)}$, whose basis states are given by:
\begin{equation}
| \widetilde{\alpha,\beta} \rangle = 
\begin{cases}
|\alpha\rangle_{1} |\beta\rangle_{2} & \text{when $\alpha=\beta$} \\
\frac{1}{\sqrt{2}} (|\alpha\rangle_{1} \otimes |\beta\rangle_{2} + |\beta\rangle_{1} \otimes |\alpha\rangle_{2}) 
& \text{when $\alpha \neq \beta$} \; .
\end{cases}
\end{equation} 
As in the previous case, we obtain the matrix elements of $\hat{S}^{A}_{1}$ and $\hat{S}^{A}_{2}$ for 
the above basis states:
\begin{equation}
\begin{split}
& \langle \widetilde{\mu,\nu} | \hat{S}^{A}_{1}| \widetilde{\alpha,\beta} \rangle \\
&= \frac{1}{2} \left\{ 
(S^{A})_{\mu\alpha}\delta_{\beta\nu} + (S^{A})_{\mu\beta}\delta_{\alpha\nu} 
+ (S^{A})_{\nu\alpha}\delta_{\beta\mu} + (S^{A})_{\nu\beta}\delta_{\alpha\mu} 
\right\}  \, , \\
& \langle \widetilde{\mu,\nu} | \hat{S}^{A}_{2}| \widetilde{\alpha,\beta} \rangle
= \langle \widetilde{\mu,\nu} | \hat{S}^{A}_{1}| \widetilde{\alpha,\beta} \rangle  \; .
\end{split}
\end{equation}
Again we have similar relations 
\begin{equation}
\begin{split}
\langle \widetilde{\mu,\nu} | \hat{S}^{A}({\tiny \yng(1,1)})  | \widetilde{\alpha,\beta} \rangle
& = \langle \widetilde{\mu,\nu} | \left\{ \hat{S}^{A}_{1} + \hat{S}^{A}_{2} \right\} | \widetilde{\alpha,\beta} \rangle \\
&= 2 \langle \widetilde{\mu,\nu} | \hat{S}^{A}_{1}| \widetilde{\alpha,\beta} \rangle
= 2 \langle \widetilde{\mu,\nu} | \hat{S}^{A}_{2}| \widetilde{\alpha,\beta} \rangle
\end{split}
\end{equation}
and we can readily see that the first-order (in $J$) effective Hamiltonian for $J_{\perp}<0$ is given again by the Heisenberg 
Hamiltonian:
\begin{equation}
\mathcal{H}_{\text{eff}}(J_{\perp}<0) 
= \frac{1}{2} (J_{\parallel} + J_{\times})  \sum_{j} S^{A}_{j}({\tiny \yng(2)}) S^{A}_{j+1}({\tiny \yng(2)})  
\label{eqn:eff-Ham-sym-ladder-FM}
\end{equation}
with the spins belonging to the symmetric 2-tensor representation ${\tiny \yng(2)}$.  
If we interpret the SU($N$) spins $\{ S_{l,j}^{A} \}$ ($l=1,2$) coming from $N$-colored fermions, this corresponds to the situation considered 
in Appendix A of Ref.~\onlinecite{Hermele-G-11}.   
Therefore, in the case of the SU($N$) two-leg ladder with identical representations for the two legs, the strong-coupling limit ($|J_{\perp}| \gg 1$) 
is described by the usual Heisenberg Hamiltonian in ${\tiny \yng(1,1)}$ or ${\tiny \yng(2)}$ regardless of the sign of $J_{\perp}$, while 
an additional $\mathcal{D}\mathcal{D}$-interaction appears in the case of $\boldsymbol{N}$-$\overline{\boldsymbol{N}}$ ladder 
[see eq.~\eqref{eqn:strong-coupling-N-Nbar}].  

The results \eqref{eqn:eff-Ham-sym-ladder-AF} and \eqref{eqn:eff-Ham-sym-ladder-FM} imply that, 
in contrast to the case of asymmetric ladder where we have several phases when $-J_{\perp} \gg J_{\parallel},J_{\times}$, 
there is only a single phase in the strong-coupling regions of the symmetric 
$\boldsymbol{N}\text{-}\boldsymbol{N}$ ladder. 
\subsection{$\mathcal{D}\mathcal{D}$ interaction}
\label{sec:DD-int}
In this Appendix, we try to express the $\mathcal{D}\mathcal{D}$-interaction appearing in Sec.~\ref{sec:strongcoupling} 
in terms of the SU($N$) spin operators.  To this end, we start from the Clebsch-Gordan decomposition of 
a product of two adjoint representations of SU($N$): 
\begin{equation}
\begin{split}
& \text{\scriptsize $N{-}1$} \left\{ 
{\scriptsize \yng(2,1,1,1) } 
\right. \otimes {\scriptsize \yng(2,1,1,1)}  \\
& \simeq   \; 
\text{\scriptsize $N{-}1$} \left\{ 
{\scriptsize  \yng(4,2,2,2) } \right.
\oplus 
\underbrace{%
\text{\scriptsize $N{-}2$} \left\{ 
{\scriptsize \yng(3,1,1) }
\right.  
\; \oplus \;
\text{\scriptsize $N{-}1$} \left\{ 
{\scriptsize \yng(3,3,2,2) }
\right.
\raisebox{1.8ex}{%
$\left.
\vphantom{{\scriptsize \yng(2,2)}}
 \right\} \text{\scriptsize $2$}
$ }
}_{\text{conjugate}} \\
&  \quad \oplus \;  \text{\scriptsize $N{-}2$} \left\{ 
{\scriptsize \yng(2,2,1) }
\right. 
\raisebox{0.9ex}{%
$\left.
\vphantom{{\scriptsize \yng(2,2)}}
 \right\} \text{\scriptsize $2$}
$ }\;
 \oplus \; 2 \times \; \text{\scriptsize $N{-}1$} \left\{  {\scriptsize \yng(2,1,1,1)}  \right. 
 \oplus  \bullet  \; .
\end{split}
\label{eqn:CG-decomp-SUN-adj-adj}
\end{equation} 
Precisely, this is valid only for $N \geq 4$ and 
the fourth representation on the right-hand side does not appear in SU(3).  
Note that there are two different adjoint representations that are conveniently distinguished by, e.g., 
the two eigenvalues $+1$ (`S') and $-1$ (`A') of the 2-site permutation operator. 
Being SU($N$)-invariant and parity-symmetric, the $\mathcal{D}\mathcal{D}$-interaction can be 
decomposed into the projectors onto the irreducible representations appearing in \eqref{eqn:CG-decomp-SUN-adj-adj}:  
\begin{equation}
\begin{split}
& \sum_{A=1}^{N^2 -1} \mathcal{D}^A_{i} \mathcal{D}^A_{i+1} =
\frac{1}{N}(N^2 - 4) P(\bullet) - \frac{4}{N} P  \left({\tiny \yng(2,1,1,1)}\right)_{\text{S}}  \\
& + \frac{1}{2N}(N^2 - 4) P  \left({\tiny \yng(2,1,1,1)}\right)_{\text{S}\oplus\text{A}} 
- \frac{1}{N}(N+2) P\left({\tiny \yng(2,2,1)}\right)   \\
& - \frac{2}{N} \left\{ P  \left({\tiny \yng(3,1,1)}\right) + P  \left({\tiny \yng(3,3,2,2)}\right)  \right\} 
+ \frac{1}{N}(N - 2) P\left({\tiny \yng(4,2,2,2)}\right)  \; . 
\end{split}
\label{eqn:decomp-DD-by-proj}
\end{equation} 
All the terms but the projector onto the symmetric (`S') adjoint (the second one) are expressed as eighth-order 
[sixth-order in SU(3)] polynomials in 
$X \equiv \mathcal{S}^A_{i} \mathcal{S}^A_{i+1}$.   
To express the projector onto the symmetric adjoint, it is convenient to introduce 
the following reflection-symmetric SU($N$)-invariant six-spin operator\cite{Roy-Q-18}
 \begin{equation}
 \mathcal{K}_{i,i+1}  \equiv d_{ABC}d_{DEF}  
 \mathcal{S}_{i}^{A} \mathcal{S}_{i}^{D} \mathcal{S}_{i}^{E} \mathcal{S}_{i+1}^{F} \mathcal{S}_{i+1}^{B} \mathcal{S}_{i+1}^{C}  \; ,
 \label{eqn:def-6spin-K}
 \end{equation}
where the SU($N$) generators $\{ \mathcal{S}_{i}^{A}\}$ are in the adjoint representation and 
the symmetric structure constants $d_{ABC}$ are defined in Eq.~\eqref{eqn:comm-rel-defining-rep}.    
Then, we can rewrite the projector with $ \mathcal{K}$ and a polynomial of $X$ as:
\begin{equation}
\begin{split}
P  \left({\tiny \yng(2,1,1,1)}\right)_{\text{S}}  =&  
\frac{1}{4}(N^2 -4) P(\bullet) 
- \frac{1}{4}(N + 2) P\left({\tiny \yng(2,2,1)}\right) \\
& + \frac{1}{4} (N - 2) P\left({\tiny \yng(4,2,2,2)}\right)
 - \frac{2}{N^2}  \mathcal{K}  \; .
\end{split}
\end{equation}
Plugging this into \eqref{eqn:decomp-DD-by-proj}, we obtain ($X = \mathcal{S}^A_{i} \mathcal{S}^A_{i+1}$):
\begin{subequations}
\begin{equation}
\begin{split}
\sum_{A=1}^{N^2 -1} \mathcal{D}^A_{i} \mathcal{D}^A_{i+1}   
& = \frac{2}{N^2} X +\frac{2(N^2+2)}{N^3}X^2  - \frac{2}{N^2} X^3  
\\ 
& \phantom{=} \;  - \frac{4}{N^3} X^4  + \frac{8}{N^3} \mathcal{K}_{i,i+1}  -\frac{2}{N}   
\end{split}
\label{eqn:DD-int-by-polynom-gen-N}
\end{equation}
for $N \geq 4$ and  
\begin{equation}
\begin{split}
& \sum_{A=1}^{8} \mathcal{D}^A_{i} \mathcal{D}^A_{i+1}   \\
&  = -\frac{4}{9}X +\frac{22}{27} X^2 +\frac{8}{27} X^3 + \frac{8}{27} \mathcal{K}_{i,i+1}  -\frac{2}{3}   
\end{split}
\end{equation}
\end{subequations}
for $N=3$.  
\section{Projection and permutation operators} 
\label{sec:peojection-permutation}
In this Appendix, we summarize some useful relations used in Sec.~\ref{sec:large-N}.  
For SU(2), it is well-known that the permutation operator that exchanges the two $S=1/2$ states is given 
simply as $\mathcal{P}(i;j)=1/2+2\mathbf{S}_{i}{\cdot}\mathbf{S}_{j}$.  The simplest way of deriving this would be to 
write $\mathcal{P}(i;j)= a + b \mathbf{S}_{i}{\cdot}\mathbf{S}_{j}$ and use that $\mathcal{P}(i;j)=1$ ($-1$) for triplet (singlet).  
For a pair of $\bolN$ or $\overline{\bolN}$ of SU($N$), we can follow the same strategy and use Eq.~\eqref{eqn:SdotS-N-N} to obtain:
\begin{equation}
\begin{split}
& \mathcal{P}(i;j) = \sum_{a,b=1}^{N} | a\rangle_{i} |b \rangle_{j} \langle b |_{i}\langle a |_{j} 
= 2 \sum_{A=1}^{N^{2}-1} S_{i}^{A} S_{j}^{A} + \frac{1}{N}   \; , \\
& \overline{\mathcal{P}}(i;j)  = \sum_{a,b=1}^{N} | \bar{a}\rangle_{i} | \bar{b} \rangle_{j} \langle \bar{b} |_{i}\langle \bar{a} |_{j}  
= 2 \sum_{A=1}^{N^{2}-1} \overline{S}_{i}^{A} \overline{S}_{j}^{A} + \frac{1}{N}    \; .
\end{split}
\label{eqn:perm-by-exch}
\end{equation}

For the pair $\bolN$ and $\overline{\bolN}$, the permutation is not defined.  Nevertheless, the projection operators 
onto the irreducible representations can be defined.  
As the pair $\bolN$ and $\overline{\bolN}$ may be decomposed as \eqref{eqn:CG-N-Nbar}, 
any projection operators can be expressed as $a + b \sum_{A=1}^{N^{2}-1}S_{i}^{A}\overline{S}_{j}^{A} $.   
Using the values in Eq.~\eqref{eqn:SdotS-N-Nbar},  we can determine the values of $a$ and $b$ to obtain:
\begin{equation}
\begin{split}
P^{(\boldsymbol{N},\overline{\boldsymbol{N}})}_{\text{singlet}}(i;j)
&= \left( \frac{1}{\sqrt{N}} \right)^{2} \sum_{a,b=1}^{N} | a\rangle_{i}| \bar{a}\rangle_{j} \langle b |_{i}\langle \bar{b} |_{j} \\
&= - \frac{2}{N} \sum_{A=1}^{N^{2}-1} S_{i}^{A}\overline{S}_{j}^{A}  + \frac{1}{N^2}     \; .
\end{split}
\label{eqn:proj-by-exch}
\end{equation}
\section{Matrix-product states}
\label{sec:MPS}
As both the SD states and the SU($N$) chiral AKLT states are correlated many-body states, 
it is convenient to express them in terms of matrix-product states.   
One (NW) of the two SD states shown in Fig.~\ref{fig:SD-CD} are expressed as: 
\begin{equation}
\begin{split}
&  \cdots \tfrac{1}{\sqrt{N}} \sum_{a=1}^{N} |a \rangle_{1\text{R}} | \bar{a} \rangle_{2,i} 
\tfrac{1}{\sqrt{N}} \sum_{b=1}^{N} | b \rangle_{1,i} |\bar{b}\rangle_{2,i+1}
\tfrac{1}{\sqrt{N}} \sum_{c=1}^{N} | c \rangle_{1,i+1} |\bar{c}\rangle_{2\text{L}} \cdots   \\
& = \sum_{a=1}^{N} \sum_{b=1}^{N} \sum_{c=1}^{N} \cdots  |a \rangle_{1\text{R}}  
\mathcal{A}^{\text{SD-NW}}(i) \mathcal{A}^{\text{SD-NW}}(i+1) 
|\bar{c}\rangle_{2\text{L}} \cdots   \; ,
\end{split}
\end{equation}
where the matrices $\mathcal{A}$ are defined by:
\begin{equation}
\left[ \mathcal{A}^{\text{SD-NW}}(i) \right]_{ab} \equiv \frac{1}{\sqrt{N}} | b \rangle_{1,i} | \bar{a} \rangle_{2,i} 
\;\; (a,b=1,\ldots,N) \; .
\label{eqn:MPS-SD-1}
\end{equation}
Similarly, we can express the other state (NE) with:
\begin{equation}
\left[ \mathcal{A}^{\text{SD-NE}}(i) \right]_{ab} \equiv \frac{1}{\sqrt{N}} |a \rangle_{1,i}  | \bar{b} \rangle_{2,i}   \; .
\label{eqn:MPS-SD-2}
\end{equation}
Two different chiral Haldane states [see Fig.~\ref{fig:SD-CD}(b)]  
can be constructed out of the preformed 8-dimensional spins (in the adjoint representation) and are given  
in the form of a product of $N{\times}N$ matrices $\mathcal{M}^{\text{AKLT-$\ell$}}$:
\begin{equation}
\otimes_{i} \mathcal{M}^{\text{AKLT-$\ell$}} (i)  \quad (\ell= \text{NE, NW}) 
\end{equation}
with 
\begin{equation}
\begin{split}
& \mathcal{M}^{\text{AKLT-NW}} (i)_{ab}
= \sqrt{\frac{2N}{N^{2}-1}} \sum_{A=1}^{N^{2}-1} (\mathcal{S}^{A})_{ab} |A\rangle_{i}     \\
&=  \frac{N}{\sqrt{N^{2}-1}} \mathcal{A}^{\text{SD-NW}}(i)_{ab} 
- \frac{1}{\sqrt{N^2 -1}} \delta_{ab} \left| \raisebox{-1.0ex}{\includegraphics[scale=0.16]{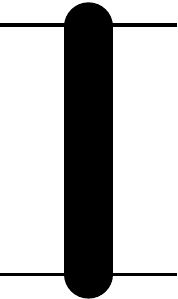}} \right\rangle_{i}    \\
& \mathcal{M}^{\text{AKLT-NE}} (i)_{ab}
= \sqrt{\frac{2N}{N^{2}-1}} \sum_{A=1}^{N^{2}-1} (\mathcal{S}^{A})^{\text{T}}_{ab} |A\rangle_{i}     \\
&=  \frac{N}{\sqrt{N^{2}-1}} \mathcal{A}^{\text{SD-NE}}(i)_{ab} 
- \frac{1}{\sqrt{N^2 -1}} \delta_{ab} \left| \raisebox{-1.0ex}{\includegraphics[scale=0.16]{./figures/RS}} \right\rangle_{i}   \; , 
\end{split} 
\label{eqn:AKLT-by-SD}
\end{equation}
where we have used Eqs.~\eqref{eqn:adj-states-by-3-3bar} and \eqref{eqn:GaGa-tensor} and 
the singlet state $\left| \raisebox{-0.6ex}{\includegraphics[scale=0.12]{./figures/RS}} \right\rangle$ 
on a rung is defined by 
\begin{equation}
\left| \raisebox{-1.0ex}{\includegraphics[scale=0.16]{./figures/RS}} \right\rangle_{i} = 
\frac{1}{\sqrt{N}} \sum_{c=1}^{N} |c \rangle_{1,i}|\bar{c}\rangle_{2,i}   \; .
\end{equation}
\section{$\mathbb{Z}_{N}{\times}\mathbb{Z}_{N}$-symmetry and string order parameters}
\label{sec:string-OP}
\subsection{Generators and order parameters}
\label{eqn:ZnxZn-gen-by-spins}
It is well-known that we can think of the SPT phases protected by on-site $\text{PSU($N$)}$-symmetry as 
protected by its discrete subgroup $\mathbb{Z}_{N}{\times}\mathbb{Z}_{N}$ \cite{Duivenvoorden-Q-ZnxZn-13,Else-B-D-13}. 
Since the string-order parameters are defined with respect to this subgroup of PSU($N$), 
we first show how we identify $\mathbb{Z}_{N}{\times}\mathbb{Z}_{N}$ in the language of SU($N$).  
For the SU($N$) generators, we use the following generalized Gell-Mann matrices, 
in which the $(N-1)$ Cartan generators $H_{k}$ $(k=1,\ldots,N-1)$ in the defining representation $\bolN$ are diagonal 
(see, e.g. chapter 13 of Ref.~\onlinecite{Georgi-book-99}):
\begin{equation}
[H_{k}(\boldsymbol{N})]_{ij} = \frac{1}{\sqrt{2 k(k+1)}} \delta_{ij} \left( \sum_{l=1}^{k} \delta_{jl} - k \delta_{j,k+1} \right)  \; .
\label{eqn:Cartan-in-GellMann}
\end{equation}
The other $N(N-1)$ off-diagonal generators are defined systematically as 
\begin{equation} 
\begin{split}
& S^{1} = M_{\text{s}}(1,2), \; S^{2} = M_{\text{a}}(1,2), \; S^{3} = H_1 , \; S^{4}= M_{\text{s}}(1,3), \\ 
& S^{5} = M_{\text{a}}(1,3) , \; S^{6} = M_{\text{s}}(2,3) , \; S^{7} = M_{\text{a}}(2,3), \; S^{8}=H_2 , \\
& S^{9} = M_{\text{s}}(1,4) , \; S^{10} = M_{\text{a}}(1,4) , \; S^{11} = M_{\text{s}}(2,4) , \\ 
& S^{12} = M_{\text{a}}(2,4) , \; S^{13} = M_{\text{s}}(3,4) , \; S^{14} = M_{\text{a}}(3,4) , \\ 
& S^{15}=H_3, \ldots
\end{split}
\end{equation}
using the following matrices:
\begin{equation}
\begin{split}
& [M_{\text{s}}(a,b) ]_{ij} = \frac{1}{2} ( \delta_{a,i}\delta_{b,j} + \delta_{a,j}\delta_{b,i} )  \\
& [M_{\text{a}}(a,b) ]_{ij} = - \frac{i}{2} ( \delta_{a,i}\delta_{b,j} - \delta_{a,j}\delta_{b,i} )  \\
&  (1 \leq a < b \leq N, \, 1 \leq i,j \leq N) \; .
\end{split}
\end{equation}

It is convenient to choose $Q$ to be diagonal, i.e.
\begin{equation}
\begin{split}
Q (\boldsymbol{N})  
& = 
\begin{pmatrix}
1 & 0 & \cdots & 0 \\
0 & \omega  & \cdots & 0 \\
\vdots & \vdots & \ddots &0 \\
0 & \cdots & 0 & \omega^{N-1} 
\end{pmatrix} \\
& \equiv \be^{- i \pi \frac{N-1}{N}} \exp \left[ i \frac{2\pi}{N}H_{\rho} (\bolN) \right] 
\;\; ( \omega \equiv \be^{i \frac{2\pi}{N}} ) \; ,
\end{split}
\label{eqn:Q-by-H}
\end{equation}
where the Cartan generator 
$H_{\rho} (\bolN)$ is expanded by the $(N-1)$ fundamental weights $\{\vec{\Lambda}_{\alpha}\}$ of SU($N$) as:
\begin{equation}
H_{\rho} = \sum_{k=1}^{N-1} \left\{ \sum_{\alpha=1}^{N-1} [\vec{\Lambda}_{\alpha}]_{k} \right\} H_{k}  \; .
\end{equation}
The reason why we chose $H_{\rho}$ this way is that the operator $Q^{N}$ with $Q=\be^{i \frac{2\pi}{N}H_{\rho}}$ 
is the identity $\mathbf{1}$ up to a phase in any irreducible representations.  

The generator $P$ in the defining representation is given by the permutation operator:
\begin{equation}
P (\boldsymbol{N}) = 
\begin{pmatrix}
0 & 0 & \cdots & 1 \\
1 & 0  & \cdots & 0 \\
\vdots & \ddots & \ddots & \vdots \\
0 & \cdots & 1 & 0
\end{pmatrix} 
\; .
\end{equation}
It is convenient to assume the following expansion similar to \eqref{eqn:Q-by-H}:
\begin{equation}
P (\boldsymbol{N}) = \be^{- i \pi \frac{N-1}{N}} \exp \left[ i \frac{2\pi}{N} G_{\rho} (\bolN) \right]  
\label{eqn:P-by-G}
\end{equation}
in which $G_{\rho}$ is expanded in the SU($N$) generators since $\text{det} \left(\be^{ i \frac{2\pi}{N} G_{\rho} (\bolN) } \right) =1$.  
We can easily see that these operators satisfy the unitarity and 
\begin{equation}
\begin{split}
& Q (\boldsymbol{N})^{N} = P (\boldsymbol{N})^{N} = 1 \, , 
\\
& Q (\boldsymbol{N}) P (\boldsymbol{N}) = \omega P (\boldsymbol{N}) Q  (\boldsymbol{N})   \; .
\end{split}
\end{equation}
These equations imply that the two $\mathbb{Z}_{N}$ are represented projectively in $\boldsymbol{N}$.  
As $Q$ and $P$ for arbitrary representations are constructed by tensoring $Q (\boldsymbol{N})$ and 
$P (\boldsymbol{N})$, it is obvious that $Q$ and $P$ are commuting when the number of boxes $n_{\text{y}}$ in the Young diagram 
is divisible by $N$.  

The expressions \eqref{eqn:Q-by-H} and \eqref{eqn:P-by-G} are convenient as once we establish them for $\boldsymbol{N}$, it is 
straightforward to generalize them to any irreducible representations.   
The generators $H_{\rho}$ and $G_{\rho}$ for $N=2$ are nothing but the $z$ and $x$ components of the spin operators,  
\cite{Kennedy-T-92-PRB}
while for $N=3$ they are given by:
\begin{equation}
H_{\rho} = 2 S^{3} \; , \;\; 
G_{\rho} = S^1 + S^4 + S^6 + \frac{1}{\sqrt{3}} \left( S^2 - S^5 + S^7 \right)
\end{equation}
with $S^{A}$ being the Gell-Mann matrices.  We can repeat the same steps to find $H_{\rho}$ and $G_{\rho}$ for 
$N\geq 4$.  

On top of the $\mathbb{Z}_{3}$ generators $P$ and $Q$, we need a set of operators $\{ X_{P},X_{Q} \}$ transforming as:
\begin{equation}
\begin{split}
& P^{\dagger} X_{P} P = \omega^{-1} X_{P} \;\; , \quad 
Q^{\dagger} X_{Q} Q = \omega X_{Q} \quad (\omega = \text{e}^{i\frac{2\pi}{N}} ) \\
& QX_{P} = X_{P}Q  \;\; , \quad 
 P X_{Q} = X_{Q}P   
\end{split}
\end{equation}
for any irreducible representations of SU($N$).   
To find $X_{Q}$, we first express $X_{Q}$ in terms of the SU($N$) generators 
and solve 
\begin{equation}
\begin{split}
& [H_{\rho}, X_{Q}^{(1)}]= X_{Q}^{(1)} \, ,\;  [H_{\rho}, X_{Q}^{(2)}]= -(N-1) X_{Q}^{(2)} \, ,  \\
& [G_{\rho}, X_{Q}^{(1)}+X_{Q}^{(2)}]= 0   \; ,
\end{split}
\end{equation}
which determines $X_Q=X_{Q}^{(1)}+X_{Q}^{(2)}$ up to an overall factor.  The other operator $X_P$ can be found 
similarly.  
For $N=2$, we reproduce the well-known results $X_Q = S^{x}$ and $X_P = S^{z}$.  
The results for $N=3$ and $4$ are a bit more complicated:
\begin{subequations}
\begin{align}
\begin{split}
& X^{N=3}_{P} = S^3 -i S^8   \\
& X^{N=3}_{Q} = \frac{1}{\sqrt{3}} \left\{
S^1+i S^2+S^4-i S^5+S^6+i S^7  \right\}  \; ,
\end{split}
\label{eqn:XpXq-Z3}
\\
\begin{split}
& X^{N=4}_{P} =  \frac{1}{3\sqrt{2}} \left\{  3 S^3 +(1-2 i) \sqrt{3} S^8 - (1+i) \sqrt{6} S^{15} \right\}    \\
& X^{N=4}_{Q} \\
&= \frac{1}{2}  \left\{ S^1 +i S^2 +S^6 +i S^7 +S^9 -i S^{10} +S^{13} +i S^{14} \right\}   \; .
\end{split}
\label{eqn:XpXq-Z4}
\end{align}
\end{subequations}
[These operators are normalized as 
 $\text{Tr}\left( X^{\dagger}_{Q}X_{Q}\right)=\text{Tr}\left( X^{\dagger}_{P}X_{P} \right) =1$ for 
the representation $\boldsymbol{N}$.]    
In fact, these expressions are valid for any representations and, to obtain $(X_P, X_Q)$ 
for a given representation $\mathcal{R}$, we just use the generators $S^{A}$ in $\mathcal{R}$.   
The expressions in Eqs.~\eqref{eqn:Q-P-Xp-Xq-SU3} and \eqref{eqn:Q-P-Xp-Xq-SU4} are obtained in this way. 
\subsection{String order parameters}
With these operators in hand, we can define the following string order parameters:
\begin{subequations}
\begin{align}
\begin{split}
& \mathcal{O}_{1}(m,n) \\
& \equiv \lim_{|i-j|\nearrow \infty} 
\Biggl\langle \left\{X_{P}(i)\right\}^{m} \left\{
\prod_{i\leq k <j} Q(k)^{n}
\right\} \left\{ X_{P}^{\dagger}(j) \right\}^{m}  \Biggr\rangle
\end{split}
\label{eqn:def-stringOP-1b} \\
\begin{split}
& \mathcal{O}_{2}(m,n) \\
& \equiv \lim_{|i-j|\nearrow \infty} 
\Biggl\langle
\left\{  X_{Q}(i) \right\}^{m}  
\left\{ \prod_{i < k \leq j} P(k)^{n}
\right\} \left\{ X_{Q}^{\dagger}(j)  \right\}^{m} \Biggr\rangle
 \; . 
 \end{split}
\label{eqn:def-stringOP-2b}
\end{align}
\end{subequations}
These operators are related to the topological classes through the selection rules.\cite{Pollmann-T-12,Hasebe-T-13} 
In fact, using the properties of (infinite-size) MPS, we can show\cite{Duivenvoorden-Q-ZnxZn-13,Tanimoto-T-15} 
that unless the projective representations corresponding 
to the two $\mathbb{Z}_{N}$ generators $P$ and $Q$ have a particular projective phase, the above string order parameters 
are {\em identically} zero.  
On the other hand, when {\em both} $\mathcal{O}_{1}(m,n)$ and $\mathcal{O}_{2}(m,n)$ are non-zero, 
the topological label $n_{\text{top}}$ must satisfy the following relation \eqref{eqn:selection-rule-SOP}:
\begin{equation}
m+n \, n_{\text{top}} \equiv 0 \quad   (\text{mod $N$})  
\end{equation}  
(the converse is not always true).  Therefore, if we make an appropriate choice of the set $\{ (m,n) \}$, 
we can identify the topological class of a given phase.  
Specifically, we take for $N=3$, 
\begin{subequations}
\begin{equation}
(m,n) = 
\begin{cases}
(1,2)  & \text{for class-1} \\
(1,1)  & \text{for class-2} 
\end{cases}
\end{equation}
and 
\begin{equation}
(m,n) = 
\begin{cases}
(1,3)  & \text{for class-1} \\
(2,1)  & \text{for class-2} \\
(1,1)  & \text{for class-3} 
\end{cases}
\end{equation}
\end{subequations}
for $N=4$.    Strictly speaking, we need to check both $\mathcal{O}_{1}(m,n)$ and $\mathcal{O}_{2}(m,n)$.   
However, for systems possessing full SU($N$) symmetry, $\mathcal{O}_{1}(m,n)=\mathcal{O}_{2}(m,n)$ is always 
guaranteed and it suffices to compute only $\mathcal{O}_{1}(m,n)$.   

It is interesting to calculate these order parameters $\mathcal{O}_{1}(m,n)$ for several model wave functions  
of the PSU($N$) SPT phases.  
The prototypes of the wave functions of the class-1 and 2 (chiral) SPT phases in SU(3) are given by 
the states AKLT-NW and AKLT-NE defined in Eq.~\eqref{eqn:AKLT-by-SD}.  For these two states, the order parameters are 
given as
\begin{equation}
\left( \mathcal{O}_{1}(1,2), \mathcal{O}_{1}(1,1) \right) =
\begin{cases}
(27/64,0) & \text{for class-1} \\
(0,27/64) & \text{for class-2} \; .
\end{cases}
\end{equation}

There are two different ways to split a single SU(4) spin in the adjoint (${\tiny \yng(2,1,1)}$) into two pieces:
\begin{subequations}
\begin{align}
& {\scriptsize \yng(1)}\otimes {\scriptsize \yng(1,1,1)} \sim \bullet \oplus {\scriptsize \yng(2,1,1)} \\
& {\scriptsize \yng(1,1)} \otimes {\scriptsize \yng(1,1)} \sim \bullet \oplus {\scriptsize \yng(2,1,1)} \oplus {\scriptsize \yng(2,2)} \; .
\end{align}
\end{subequations}
If we apply the valence-bond (or AKLT) construction to the above two decompositions, we can construct 
the model MPSs for the chiral $(\mathbf{4},\bar{\mathbf{4}})$ (class-3) [or $(\bar{\mathbf{4}},\mathbf{4})$; class-1] and 
the non-chiral $(\mathbf{6},\mathbf{6})$ (class-2), respectively.  The values of SOP for these model states are:
\begin{equation}
\begin{split}
& \left( \mathcal{O}_{1}(1,3), \mathcal{O}_{1}(2,1), \mathcal{O}_{1}(1,1) \right)  \\
& = 
\begin{cases}
((8/15)^2,0,0) & \text{for class-1} \\
(0,(2/15)^2,0) & \text{for class-2} \\
(0,0,(8/15)^2) & \text{for class-3} \; .
\end{cases}
\end{split}
\end{equation}
\section{Non-Abelian bosonization and SU($\boldsymbol{N}$)  spin chain}
\label{Appbosonization}
In this Appendix, we review in a nutshell the continuum description of the  SU($N$) spin operator
of Eq. (\ref{spinop}) in terms of the currents $J^{A}_{\text{R,L}}$ and the WZNW primary field $g$ of the SU($N$)$_1$ CFT.
As is well-known, the low-energy properties of the SU($N$) Sutherland model can be extracted from the one-dimensional U($N$) Hubbard chain in the large repulsive $U$ limit for a $1/N$-filling with Fermi-wavector $k_{\text{F}} = \pi/(N a_0)$ with Hamiltonian:
\begin{equation}
\begin{split}
\mathcal{H}_{\text{Hubbard}} = & - t \sum_i \sum_{\alpha=1}^{N} \left( c^{\dagger}_{\alpha,i+1} c_{\alpha,i} + \text{H.c.} \right)   \\
& + \frac{U}{2} \sum_{i, \alpha,\beta} n_{\alpha,i}   n_{\beta,i} \left( 1 - \delta_{\alpha\beta} \right) \;  ,
\end{split}
\label{hubbardSUN}
\end{equation}
where $c^{\dagger}_{\alpha,i}$ creates a spinless fermion in the band $\alpha=1, \ldots N$ of the site $i$ and $
n_{\alpha,i} = c^{\dagger}_{\alpha,i} c_{\alpha,i}$ is the occupation number. 
When the Hubbard interaction $U$ is sufficiently large, the system becomes a Mott insulator and 
the charge degrees of freedom gets decoupled from the low-energy physics.  The physical properties of this
Mott-insulating phase are described by the SU($N$) Sutherland model:\cite{Affleck-NP86,Affleck-88,James-K-L-R-T-18,Itoi-K-97,Assaraf-A-C-L-99,Manmana-H-C-F-R-11}
\begin{equation}
{\cal H}_{\text{Sutherland}} = J   \sum_i \sum_{A=1}^{N^2-1} \hat{S}^{A}_{i} \hat{S}^{A}_{i+1} ,
\label{Sutherland-75app}
\end{equation}
where $J= 4 t^2/U$ is the spin-exchange and $\hat{S}^{A}_i = c^{\dagger}_{\alpha,i}  [S^{A}]_{\alpha \beta}   c_{\beta,i} $
[$\mbox{Tr} (S^A S^B) = \delta^{AB}/2$] is the SU($N$) spin operator on site $i$.  
To guarantee the defining representation $\boldsymbol{N}$ at each site, we impose the local constraint: 
$\sum_{\alpha} n_{\alpha,i}=1$ ($\forall i$).   
The model (\ref{Sutherland-75app}) is integrable and has $N-1$ gapless modes that are described by the SU($N$)$_1$ CFT
with central charge $c=N-1$. A continuum description of the SU($N$) spin operator can be derived from the continuum
expression of the lattice fermion in terms of $N$ chiral Dirac fermions: 
\begin{equation}
c_{\alpha,i}/\sqrt{a_0} \rightarrow e^{-i k_{\text{F}} x } L_{\alpha} (x)+ e^{i k_{\text{F}} x }  R_{\alpha}(x) \;  ,
\label{eqn:exp-Dirac}
\end{equation}
which enables us to obtain:
\begin{equation}
\hat{S}^{A}_i/ a_0 
\sim J^{A}_{\text{R}} + J^{A}_{\text{L}} + e^{i 2k_{\text{F}} x }  L^{\dagger}_{\alpha} [S^{A}]_{\alpha \beta}  R_{\beta} 
+ \text{H.c.} \; ,
\label{spinopapp}
\end{equation}
where $J^{A}_{\text{L}} = L^{\dagger}_{\alpha} [S^{A}]_{\alpha \beta}  L_{\beta}$ is the left SU($N$)$_1$ current  
with the right one defined similarly. The 2$k_{\text{F}}$ SU($N$) spin density $N^{A}$ of Eq. (\ref{2kf}) is obtained 
by taking the average over the fully gapped charge mode: 
$N^{A} = \left\langle  L^{\dagger}_{\alpha} [S^{A}]_{\alpha \beta}  R_{\beta} \right\rangle_{\text{c}}$.   
We then introduce $N$ left-right moving bosons $\varphi_{\alpha \text{L,R}}$  
to bosonize the Dirac fermions \cite{James-K-L-R-T-18,Gogolin-N-T-book} :
\begin{eqnarray}
 L_{\alpha} = \frac{\kappa_{\alpha}}{\sqrt{2 \pi a_0}} \; e^{ - i \sqrt{4 \pi}\varphi_{\alpha \text{L}}}, \;  
 R_{\alpha} = \frac{\kappa_{\alpha}}{\sqrt{2 \pi a_0}} \; e^{ i \sqrt{4 \pi}\varphi_{\alpha \text{R}}}  ,
\label{bosoabeleq}
\end{eqnarray}
where $[\varphi_{\alpha \text{R}}, \varphi_{\beta \text{L}} ] = i \delta_{\alpha\beta}/4$ and $\kappa_{\alpha}$ are Klein factors to ensure 
the anticommutation of fermions with different colors: $\{\kappa_{\alpha}, \kappa_{\beta} \} = 2   \delta_{\alpha\beta} $,
$\kappa^{\dagger}_{\alpha} = \kappa_{\alpha}$. The 2$k_{\text{F}}$ SU($N$) spin density  can thus be expressed in terms
of these bosonic fields as
\begin{equation}
N^{A} =  \frac{\kappa_{\alpha} \kappa_{\beta} i^{\delta_{\alpha\beta}}}{2 \pi a_0}  [S^{A}]_{\alpha \beta}    
\left\langle e^{  i \sqrt{4 \pi}\varphi_{\alpha \text{L}} + i \sqrt{4 \pi}\varphi_{\beta \text{R}} } \right\rangle_{\text{c}} 
\label{BoseNA}
\end{equation}
after averaging over the gapped charge sector ($\langle \cdots \rangle_{\text{c}}$).  
The average in Eq.~\eqref{BoseNA} can be conveniently done by switching to a new basis where the charge degrees 
of freedom are separated.  
Specifically, we introduce a charge boson field $\Phi^{\text{c}}$ and 
$N-1$ ``spin'' [i.e., SU($N$)] fields $\Phi^{\text{s}}_{m}$ ($m=1, \ldots N-1$) through the following orthogonal 
transformation:\cite{Banks-H-N-76,Assaraf-A-C-L-99}
\begin{equation}
\begin{split}
&\Phi^{\text{c}}_{\text{R,L}} = \frac{1}{\sqrt{N}} \sum_{\alpha =1}^{N} \varphi_{\alpha \text{R,L} }   \\
&\Phi^{\text{s}}_{m \text{R,L}} = \sum_{p=1}^{N-1}  [H_{m}(\boldsymbol{N})]_{pp} \varphi_{p\text{R,L}}  \; ,
\end{split}
\label{SUNbasis}
\end{equation}
where the diagonal Cartan generators $\{H_{m}(\boldsymbol{N})\}$ are defined in \eqref{eqn:Cartan-in-GellMann} 
The inverse transformation can be written in a compact way as:
\begin{equation}
\begin{split}
 \varphi_{\alpha \text{R,L} }  &= \frac{\Phi^{\text{c}}_{\text{R,L}}}{\sqrt{N}} 
 + \sum_{m=1}^{N-1}  [\vec{\omega}_{\alpha}]^{m} \Phi^{\text{s}}_{m \text{R,L}}    \\
 & \equiv   \frac{\Phi^{\text{c}}_{\text{R,L}}}{\sqrt{N}} + {\vec  \omega}_{\alpha} \cdot {\vec  \Phi}^{\text{s}}_{\text{R,L}}  ,
 \end{split}
\label{invSUN}
\end{equation}
where ${\vec  \omega}_{\alpha}/\sqrt{2}$ ($\alpha=1,\ldots,N$) 
are the $(N-1)$-dimensional weight vectors 
in the fundamental representation $\boldsymbol{N}$ of the SU($N$) which satisfy the following simple relations:
\begin{subequations}
\begin{align}
& \sum_{\alpha=1}^N \vec{\omega}_\alpha = \vec{0}, \\
& \sum_{\alpha=1}^N [\vec{\omega}_\alpha]^m [\vec{\omega}_\alpha]^{m'} = \delta_{mm'}, \\
& \vec{\omega}_{\alpha} {\cdot} \vec{\omega}_{\beta}
= \delta_{\alpha \beta} -\frac{1}{N}. 
\end{align}
\label{weightSUN}
\end{subequations}
For instance, in the $N=3$ case, we have:
\begin{equation}
{\vec \omega}_1 = \begin{pmatrix} \frac{1}{\sqrt{2}},  \frac{1}{\sqrt{6}} \end{pmatrix} \, , \; 
{\vec \omega}_2 = \begin{pmatrix} -\frac{1}{\sqrt{2}},  \frac{1}{\sqrt{6}} \end{pmatrix} \, , \; 
{\vec \omega}_3 = \begin{pmatrix} 0,  -\sqrt{\frac{2}{3}}  \end{pmatrix} \; .
\label{weightSU3}
\end{equation}
The Hamiltonian of the Sutherland model in the far infrared regime neglecting the marginal irrelevant current-current
interaction takes then a simple form in terms of these spin bosonic fields:
\begin{eqnarray}
 {\cal H}_{\text{Sutherland}} & =& v \int dx \sum_{m=1}^{N-1}  \left\{ \left(\partial_x  \Phi^{\text{s}}_{m \text{L}}\right)^2 + 
\left(\partial_x  \Phi^{\text{s}}_{m\text{R}}\right)^2  \right\} \nonumber \\
&=& v \int dx \left\{ \left(\partial_x  {\vec \Phi}^{\text{s}}_{\text{L}}\right)^2 + 
\left(\partial_x   {\vec \Phi}^{\text{s}}_{\text{R}}\right)^2  \right\}   \; .
\label{SutherlandHambose}
\end{eqnarray} 
A free-field representation of the 2$k_{\text{F}}$ SU($N$) spin density can also be obtained from 
the identity (\ref{2kf}) $N^{A}  =   i  \lambda   \; {\rm Tr} ( g  S^A)$ and \eqref{BoseNA}:
\begin{subequations}
\begin{align}
& \lambda = \frac{\sqrt{N}}{2 \pi a_0^{1/N}} \left\langle e^{  i \sqrt{4\pi/N} \Phi_{\text{c}}} \right\rangle_{\text{c}}  \label{lambda} \\
& g_{\beta \alpha} = \frac{\kappa_{\alpha} \kappa_{\beta} i^{\delta_{\alpha\beta}-1}}{\sqrt{N}}
: e^{  i \sqrt{4 \pi} {\vec  \omega}_{\alpha} \cdot {\vec  \Phi}^{\text{s}}_{\text{L}}
+ i \sqrt{4 \pi}  {\vec  \omega}_{\beta} \cdot {\vec  \Phi}^{\text{s}}_{\text{R}}  }:  \; ,
\end{align}
\label{freefieldrepWZWGfield}
\end{subequations}
where $ \Phi^{\text{c}} \equiv  \Phi^{\text{c}}_{\text{R}} +  \Phi^{\text{c}}_{\text{L}} $.  
A more rigorous free-field representation of the SU($N$)$_1$ WZNW $g$ primary field was obtained
in Ref. \onlinecite{Fuji-L-17}  (see Appendix B) where the Klein factors were constructed out of the zero mode operators of the
spin bosonic fields.  

We now argue that the non-universal constant $\lambda$ in Eq.~\eqref{freefieldrepWZWGfield}, which
results from the average over the fully gapped charge sector, can be chosen real.   To this end, we need to determine the 
umklapp term which opens the mass gap in the charge sector and pins the charge field $\Phi_c$.   
As is discussed in Ref.~\onlinecite{Assaraf-A-C-L-99},
when $N \ge 2$, this term is generated at higher orders in perturbation theory. 
As the fermion operators are expanded, at low energies, as the Hubbard Hamiltonian \eqref{hubbardSUN} 
consisting of the lattice fermion operators $c_{\alpha,i}$ \eqref{eqn:exp-Dirac}
has the following expansion in the momentum change $q$:
\begin{equation}
{\cal H}_{\text{Hubbard}} \simeq {\cal H}_{q=0} + {\cal H}_{q= \pm 2k_{\text{F}}} + {\cal H}_{q= \pm 4k_{\text{F}}} + \ldots \; .
\label{hubbardecomp}
\end{equation}
Since $k_{\text{F}}  = \pi/(Na_0)$, the umklapp terms with $q=\pm 2N k_{\text{F}}$ give a uniform (i.e., non-oscillatory) 
contribution which can open a gap in the charge sector. The umklapp operator which is an SU($N$) singlet 
and translation invariant with the lowest scaling dimension is:
\begin{equation}
\prod_{\alpha=1}^{N} L^{\dagger}_{\alpha} R_{\alpha} + \text{H.c.}  \; .
\label{umklappterm}
\end{equation}
Its generation depends on the parity of $N$. When $N=2p$, the operator (\ref{umklappterm}) appears 
from $p-1$ operator product expansions (OPE) between two $ {\cal H}_{q= \pm 4k_{\text{F}}}$ terms so that:
\begin{equation}
\begin{split}
{\cal H}^{N=2p}_{\text{umklapp}} &\sim  (-1)^{p-1} U^{p} \prod_{\alpha=1}^{N} L^{\dagger}_{\alpha} R_{\alpha} + \text{H.c.}  \\
&\sim - U^{p} \cos\left(\sqrt{4\pi N} \Phi_{\text{c}}  \right) ,
\end{split}
\label{umklapptermeven}
\end{equation}
where we have used the identification (\ref{bosoabeleq}) and the definition (\ref{SUNbasis}) of the charge field $\Phi_{\text{c}}$.
The sine-Gordon potential (\ref{umklappterm}) leads us to conclude that the charge field
is locked on configurations such that $\langle \Phi_{\text{c}} \rangle = k \sqrt{ \pi/N}$ ($k$: integer).  
One can freely choose the ground state of the sine-Gordon umklapp potential such that $\langle \Phi_{\text{c}} \rangle = 0$
so that $\lambda$ in Eq.  (\ref{lambda})  can be taken real.
When $N=2p+1$ is odd, we need $p-1$ OPE between two $ {\cal H}_{q= \pm 4k_{\text{F}}}$ terms and 
then a last OPE with ${\cal H}_{q= \pm 2k_{\text{F}}}$ to produce the umklapp term (\ref{umklappterm}).
We thus find: 
\begin{equation}
\begin{split}
{\cal H}^{N=2p+1}_{\text{umklapp}} &\sim  i (-1)^{p}  U^{p+1} \prod_{\alpha=1}^{N} L^{\dagger}_{\alpha} R_{\alpha} + \text{H.c.}  \\
&\sim - U^{p+1} \cos\left(\sqrt{4\pi N} \Phi_{\text{c}} \right) \; ,
\end{split}
\label{umklapptermodd}
\end{equation}
from which we conclude as in the even $N$ case that $\lambda$ is real.

\include{draft.bbl}

\end{document}

%% file: draft.bbl
%